\documentclass[useAMS,usenatbib,asm]{mn2e}         
\usepackage{graphicx,fleqn,color}
\usepackage{natbib}
\usepackage{amsmath}
\usepackage{amssymb}
\usepackage{moreverb}
\usepackage{float}

\begin{document}

\date{Accepted. Received ; in original form }
\pagerange{\pageref{firstpage}--\pageref{lastpage}} \pubyear{2014}

\title[Characterizing the Galactic warp with Gaia]{Characterizing the Galactic warp with Gaia:
 I. The tilted ring model with a twist } 
\author[H. Abedi, C. Mateu, L.A. Aguilar, F. Figueras \& Romero-G\'omez, M.]{Hoda Abedi$^1$\thanks{e-mail: habedi@am.ub.es}, 
Cecilia Mateu$^{2,3}$, Luis A. Aguilar$^{2}$, Francesca Figueras$^1$ \newauthor and Merc\`e Romero-G\'omez$^1$\\ \\
$^{1}$ Dept. d'Astronomia i Meteorologia, Institut de Ci\`encies del Cosmos, Universitat de Barcelona, IEEC,\\
 Mart\'{\i} i Franqu\`es 1, E08028 Barcelona, Spain \\
$^{2}$ Instituto de Astronomia, UNAM, Apartado Postal 877, 22860 Ensenada, B.C., Mexico\\
$^{3}$ Centro de Investigaciones de Astronom\'{\i}a, AP 264, M\'erida 5101-A, Venezuela}

\maketitle
\label{firstpage}

\begin{abstract}
We explore the possibility of detecting and characterizing the warp 
of the stellar disc of our Galaxy using synthetic Gaia data. The 
availability of proper motions and, for the brightest stars radial 
velocities, adds a new dimension to this study. A family of Great 
Circle Cell Counts (GC3) methods is used. They are ideally suited 
to find the tilt and twist of a collection of rings, which allow us 
to detect and measure the warp parameters. To test them, we use 
random realizations of test particles which evolve in a realistic 
Galactic potential warped adiabatically to various final 
configurations. In some cases a twist is introduced additionally. 
The Gaia selection function, its errors model and a realistic 3D 
extinction map are applied to mimic three tracer populations: OB, A 
and Red Clump stars. We show how the use of kinematics improves the 
accuracy in the recovery of the warp parameters. The OB stars are 
demonstrated to be the best tracers  determining the tilt angle with 
accuracy better than $\sim0.5$ up to Galactocentric distance of 
$\sim16$ kpc. Using data with good astrometric quality, the same 
accuracy is obtained for A type stars up to  $\sim13$ kpc and for Red
Clump up to the expected stellar cut-off. Using OB stars the twist 
angle is recovered to within $<3\degr$ for all distances.

\end{abstract}


%====================================================

\section{Introduction}
It is widely accepted that warps of disc galaxies are a common phenomena (as common as spiral structure), yet they are still not fully 
understood (see \cite{garcia} for a historical review). From observational studies in external galaxies, Briggs's
 laws \citep[]{briggs90} state that discs are generally flat inside the $R_{25}$ radius that is inside the solar Galactocentric radius in
 our Milky Way (MW);  
the line of nodes is straight out to $R_{26.5}$, indicating that the self-gravity of the disc is important; and, farther out the line of 
nodes advances in the direction of galactic rotation indicating that the warps are not quite in equilibrium at large radii. Note 
that $R_{25}$ and $R_{26.5}$ are the radius of galaxy to the respectively $25$ $\mathrm{mag\, arcsec}^{-2}$ and $26.5$ $\mathrm{mag\, arcsec}^{-2}$ isophotes. 
As discussed by \cite{cox96}, as the stellar warps usually follow the same warped surface as
 do the gaseous ones, there is strong evidence that warps are mainly a gravitational phenomenon. Furthermore, as \cite{sellwood13} 
stated, the ubiquity of warps in external galaxies suggests that they are either, repeatedly regenerated, or a long-lived phenomenon.
 In any case, warped discs represent a theoretical challenge and, if properly understood, can be a valuable probe into the mass
 distribution in the outer disc and the halo in its vicinity \citep[]{binneyR92}.

From the time when the first 21-cm observations of our Galaxy became available, the large-scale warp in the HI gas disc  
became apparent \citep[][ among others]{burke,Westerhout,oort58}. More than fifty years later, \cite{levine}  
has re-examined the outer HI distribution proposing that the warp of gas is well described by two Fourier modes, the m=2 mode accounting
 for a large asymmetry between the northern ($l=90\degr$) and southern warps ($l=270\degr$). Meanwhile, \cite{reyle}, from the dust and stars
 distribution, using 2MASS infrared data, found the stellar component, in a first approximation, to be well modelled by an S-shaped warp with
 a significantly smaller slope that the one seen in the HI warp. Several authors have tried to estimate the phase angle 
of the line of nodes with
respect to the Sun-Galactic centre line. Values range between $\sim-5\degr$ \citep[]{lopez02} and $\sim15\degr$ \citep[]{momany}.
 These morphological studies of the MW warp do not allow us, at present, to disentangle which
 are the mechanisms that are able to explain it.

Many efforts have been directed toward understanding warps on a theoretical basis and, at least, three mechanisms have been
 proposed for their existence \citep[see the excellent reviews by][]{binney98,sellwood13}. One of these mechanisms posits that warps
 are free normal modes of oscillation of the galactic disc. \cite{lynden65} suggested that warps could result from 
a persisting misalignment between the spin axis and the disc normal, a suggestion that was later elaborated by \cite{toomre}. From 
these pioneer works, the bending modes have long been suspected as the mechanism creating and maintaining warps. The distribution 
of matter in the halo would control the ability of the disc to sustain a long-lived bending wave. In this context, \cite{sparcas88}
 proposed that an axisymmetric, but flattened halo, could not wind up into corrugation waves but continue, indefinitely as a
 standing wave in the disc. However, it was found that a careful arrangement in the mass distribution (shape and density profile)
 is needed to obtain these long-lived modes, making this an unlikely scenario. On top of this, it turns out that the response of
 the halo to the warped disc distribution, which was not taken into account in the original mode calculations, invalidates this
 approach. \cite{binney98} stated that none of these mechanisms is viable, the only surviving explanation is that the warps
 are driven by the accretion of material \citep[also discussed in][]{lopez02b,olano}, that is, by the accretion
 of angular momentum. The tidal interaction between galaxies (forcing by satellites) has also been proposed to produce asymmetrical
 warps. As an example, \cite{levine} and \cite{wb06} proposed that a passage of the Large Magellanic Cloud could have generated the
 Galactic warp. Nowadays it seems that these satellite tides are generally too weak to produce warps of the amplitude observed.

The first kinematic warp analysis were inferred from Hipparcos proper motions of OB type stars \citep[]{drimmel0}.
 These authors concluded that the kinematics observed toward the anticentre were inconsistent with the ones expected for a long-lived warp, 
showing that only a very high bias in the photometric distances and/or a high warp precession rate could explain the observations.
Later, \cite{bobylev10}, using the Tycho-2 kinematic data of Red Clump (RC) stars, associated the observed rotation of the stellar 
system around the Sun-Galactic centre axis to the Galactic warp. However, the derived angular velocity obtained is opposite to
 the values previously obtained by \cite{miyamoto98}. More recently, \cite{bobylev13} found that the available stellar proper 
motion samples alone do not allow complete information to be obtained. Nevertheless, he undertook a first analysis 
using six-dimensional phase-space data for a small Cepheids 
sample. Discrepant results obtained from all these works demonstrate the difficulty, at present, to disentangle the kinematic 
signature of the warp from other nearby and local perturbations.

At present, what is needed is better information that could help us to
disentangle among the various competing scenarios. The warp of our
own Galaxy is the one closest to us and thus, potentially a lot of detailed
information may be gleaned from it. At the dawn of the Gaia era, a whole
hitherto  unexplored dimension opens up: adding good kinematical information
of in situ stars partaking in the warp. This dimension must be explored:
To what extent is it that Gaia data will be able to characterize the stellar
disc warp and up to what distance? To answer this question, new detection
and characterization tools must be devised and tested with Gaia mock
catalogues and their limits identified.

In this paper we do not develop a fully self-consistent simulation of a warp, but rather a first simplified kinematic model for our MW warp.
This is because it is not our goal to dwell into dynamical scenarios, but rather have a reasonable toy model with which to asses the
 real possibilities of Gaia in detecting and characterizing the warp.
 In section~\ref{sec:warpmodel} we present the warp models used. In section~\ref{sec:samples} we describe the generation of initial
 conditions, the selected tracer populations and the simulation of observational errors, extinction and the Gaia selection function.
 In section~\ref{sec:warpdetect} we present the family of GC3 methods and the method used to compute
 the warp twist and tilt angles. In section~\ref{sec:results} we discuss our results for the different GC3 methods, tracer populations 
and warp models. Our conclusions and expectations for the future, when the real Gaia database becomes available are 
presented in section~\ref{sec:conclusions}.
%====================================================
\section{The warp model}
\label{sec:warpmodel}
We model the axisymmetric part of the MW's potential following Allen $\&$ Santillan (A$\&$S) \citep[]{AS}. This 3D potential model consists 
of a spherical bulge, a Miyamoto-Nagai disc \citep[]{MN1975} and a massive spherical halo. The rotation curve of this potential model
 follows the one of MW. The adopted observational constraints of the model are summarized in table 1 of \cite{AS}. 
The total axisymmetric mass is $M_T = 9 \times 10^{11}\, M_{\odot}$. This is in a good agreement with the recent observational 
value which is  $M_T = 10^{+3}_{-2} \times 10^{11}\, M_{\odot}$ \citep[]{xue08}.

A first element required to model a galactic disc warp is a transformation that can be applied to 
an initially flat potential function or particle configuration, and distort it according to a specific warp model. In this section we 
introduce two non-lopsided warp models: the untwisted warp model in which the warp is applied directly to the disk potential; and the
 twisted warp model, where we twist the phase-space coordinates of our particles. In this paper,
 the line of nodes is defined to coincide with the X-axis, which goes along the Sun-Galactic centre direction;
the Y-axis perpendicular to it, positive
 in counter-clockwise direction as seen from the North Galactic Pole; and the Z-axis perpendicular to the flat Galactic plane. 
\subsection{The untwisted warp model}
\label{subsec:warp}
\begin{figure}
\begin{center}
\includegraphics[width=60mm, angle=270]{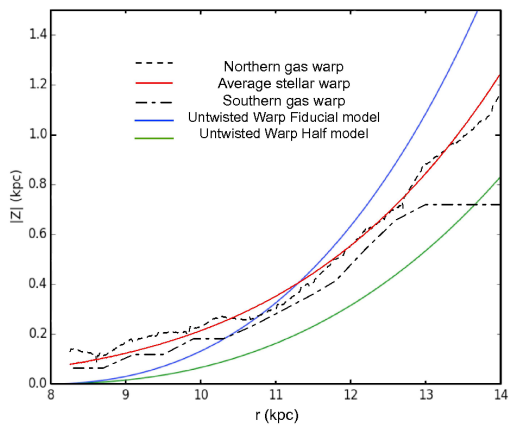}
 \caption{Maximum amplitude of the vertical height of the warp as a function of the Galactocentric radius for the full and half warp models 
respectively in  blue and green. Black dashed and dot-dashed lines and the solid red line show results from observations as 
indicated in the legend \citep[]{lopez02}.}
 \label{fig:warp_model_compare}                              
\end{center}
\end{figure}

 As a first warp model, we consider a non-precessing warp with a straight line of nodes, i.e. with null twist
 angle. To accomplish this, we rotate the Cartesian 
coordinates around the X-axis with tilt angle $\psi$ that is a function of the Galactocentric spherical $r$ coordinate of each point:
\begin{equation}\label{1}
\psi (r; r_1 ,r_2 , \psi _2, \alpha) = \begin{cases}
0, & r \leqslant  r_{1} \\
\psi _{2}(\frac{r-r_1}{r_2 - r_1})^{\alpha},  & r > r_{1}
\end{cases}
\end{equation}
Where $r_1$ and $r_2$ are the Galactocentric radii where the warp begins and finishes respectively, which are chosen to be at  $r_1=\, 8$ 
kpc and $r_2=\, 20$ kpc. The resulting warp has a tilt angle increasing as a power law, whose exponent is $\alpha$ and such that 
at $r_2$ it has a value equal to $\psi_{2}$. From observations we know that at a Galactocentric distance of $12$ kpc, the maximum height of 
the warp is about $630$ pc \citep[]{lopez02}, which corresponds to a maximum tilt angle of $\sim3^ \circ$ at $12$ kpc. By fitting our warp model to 
these values, we obtain:  $\alpha=\, 2$ and $\psi_{2}=\, 27^\circ$. This model is hereafter called the Untwisted Warp Fiducial model (UWF). 
In Figure \ref{fig:warp_model_compare}, the maximum amplitude of the warp with respect to Galactocentric radius is plotted. 
Our UWF model is overestimating the values obtained from the observations at large radii. 
So we introduce another warp model of the same properties but with $\psi_{2}=\, 13.5^\circ$ which we will call hereafter as
 the Untwisted Warp Half model (UWH). These two different warps will model two extremes of the MW's actual warp. In what follows  
 will use these untwisted models to warp the potential. 
\subsection{The twisted warp model}
\label{subsec:twist}
In order to model more complex shapes for the Galactic warp, we introduce a warp model with a twisted line of nodes.
 In this model we take the test particles that are integrated and relaxed in the warped potential modelled with the UWF and 
rotate their phase-space coordinates with respect to the Z-axis. Note that, this
 twist model is used to twist the phase-space coordinates of the particles. Here we do not apply this model to the potential, 
because, if we twist the line of nodes of our already warped potential, we do not expect the particles to be able to follow the potential
 and reach statistical equilibrium with it. The rotation is done using the following twist angle:
\begin{equation}\label{2}
\phi (r; r_1 ,r_2 , \phi _{max}) = \begin{cases}
0, & r \leqslant  r_{1} \\
\phi _{max}(\frac{r-r_1}{r_2 - r_1})^2,  & r > r_{1}
\end{cases}
\end{equation}
As for the tilt, the twist begins at $r_1=8$ kpc and finishes at $r_2=20$ kpc. 
We consider two twist models in this paper; one with $\phi _{max}=20^\circ$ and other with $\phi _{max}=60^\circ$ which hereafter will
be respectively called as TW1 and TW2 models. These values for $\phi _{max}$ are chosen to test Gaia capabilities to measure the twist. No observational 
constrains for this parameter for MW's gas or the stellar component is available at present. From external galaxies we know  
\citep[according to][third rule]{briggs90} that at large radii the line of nodes measured in the plane of the inner galaxy advances significantly in
 the direction of galaxy rotation for successively larger radii, so the line of nodes forms a loosely wrapped leading spiral.
\section{Building the warped samples}
\label{sec:samples}
\subsection{The Initial conditions}
\label{subsec:theosamp}
We generated sets of initial conditions that follow the density distribution of a Miyamoto-Nagai disc using 
the parameters
 of the Allen $\&$ Santillan Galactic model. This is done using the Hernquist method \citep[]{hernquist}. The velocity
 field is approximated using the first order moments of the collision-less Boltzmann equation simplified with the epicyclic approximation. 
The asymmetric drift is taken into account in the computation of the tangential components of the velocities.
We generated test particles for three different stellar populations: RC K-giant stars, and main sequence 
A and OB type stars. 
For this, we assign the corresponding scale height and velocity dispersions at the Sun's position to each 
test particle sample (see Table \ref{tab1}). The total number of stars for each 
tracer are shown in Table \ref{tab2}. These are calculated in such a way that the number
of stars of each tracer in a cylinder of radius 100 pc centred on the Sun position
is normalized to the number found in this cylinder
using the new Besan\c{c}on Galaxy Model \citep[]{czekaj}.
We locate the Sun at Galactocentric cartesian coordinates $(-8.5,0,0)$ kpc. We adopt a circular velocity for the Local Standard of Rest (LSR) 
 of $V_c(R_\odot)=220$ $\mathrm{kms}^{-1}$ which is consistent with both; our imposed potential model, i.e. A$\&$S model, and the observational 
constrains \citep[]{mcmillan}. Also, we consider a peculiar velocity of the Sun with respect to the LSR of
 $(U,V,W)_{\odot}=(11.1, 12.24, 7.25)$ $\mathrm{kms}^{-1}$ \citep{schonrich10}. 

The Galactic warp is a feature of the outer parts of the Galactic disc. In order to avoid integrating the path of particles 
that will not be of use later, after  
generating the initial conditions, we discard the ones whose apocentre radius is smaller than 8 kpc. This is done using the Lindblad
 diagram as explained by \cite{aguilar08}. In Table \ref{tab2} the total number of stars satisfying this condition is presented.
\begin{table}
\caption{The velocity dispersions of different stellar tracers at the Solar neighbourhood and the corresponding disc scale heights \citep[]{aumer09}.}
\label{tab1}
\begin{tabular}{@{}lccc}
\hline
Tracer & RC & A & OB\\
\hline
$\sigma_U$ [km/s] & 30 & 15 & 10\\
$\sigma_W$ [km/s] & 16 & 9 & 6\\
$z_d$ [pc] & 300 & 100 & 50 \\
\hline
\end{tabular}
\end{table}
\begin{table}
\caption{The surface number density ($\Sigma$) and the total number of stars of each tracer. Number of stars outside the 
lindblad hole refers to the number of stars with apocentre radius larger than 8 kpc.}
\label{tab2}
\begin{tabular}{@{}lccc}
\hline
Tracer & RC & A & OB\\
\hline
$\Sigma$ $[stars/pc^2]$ & 0.056 & 0.048 & 0.003\\
Number of stars & $57\times 10^6$ & $48\times 10^6$ & $3.2\times 10^6$\\
Number of stars outside& $36\times 10^6$ & $30\times 10^6$ & 1.8$\times 10^6$\\
 the Lindblad hole& & & \\
\hline
\end{tabular}
\end{table}
\subsection{Relaxation and steady state}
\label{subsec:relax}
\begin{figure*}
\begin{center}
\includegraphics[width=42mm, angle= 270]{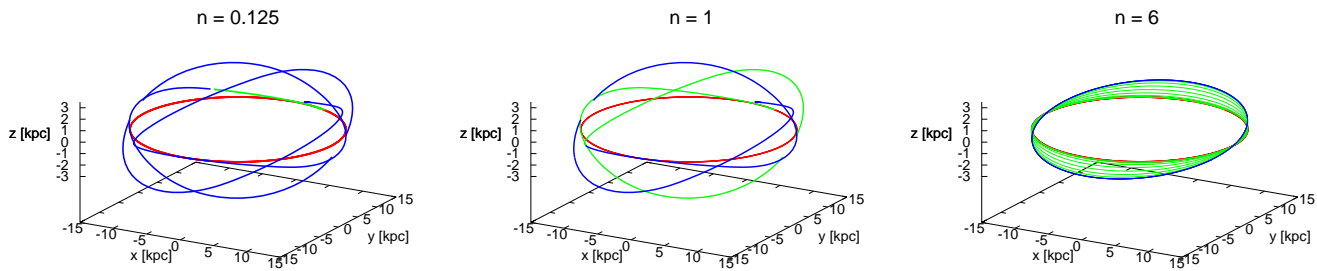}
 \caption{ The orbit of a star with an initial circular orbit with a Galactocentric radius of 14 kpc integrated for $t_1 \,=\, 2P$ in the A$\&$S potential 
 with a flat Miyamoto-Nagai disc (in red), then integrated for $t_{grow} \,= \, n \times P$ while the disc potential is gradually being warped as a function of time (see Equation
\ref{timedepeq}) (in green) and finally integrated for $t_2 \,=\, 2P$ in the final state of the warped potential (in blue). Each panel shows the orbits for 
different values of $n$. Note that P is the orbital period of a star with a circular orbit in the A$\&$S potential
 located at 20 kpc. It is clear that as we increase the value of $n$, i.e. warping the potential slower, the star can follow the potential
more closely and acquire less motion orthogonal to the instantaneous plane of the warped disc potential. Choosing $n=6$, the potential
is being warped adiabatically enough that the star can keep its circular orbit within the warped plane. }
 \label{fig:orbit}                              
\end{center}
\end{figure*}

We apply our untwisted warp models to the Miyamoto-Nagai disc potential of A$\&$S.
In order to be able to integrate the test particles in this warped potential, we calculate the corresponding warped forces (See Appendix A for
 details).\\
Starting from a set of test particles that are relaxed in the A$\&$S potential, if we abruptly bend the disc potential, 
the particles will loose their near-circular orbits and increase their velocity dispersion. To avoid this, we should warp the potential 
adiabatically, in other words, we should do it slowly enough, so that the particles can follow the bended potential and not be left behind. 
To do this, we make $\psi_2$ a function that grows with time and reaches its maximum value $\psi_{max}$ at 
time $t_{grow}$. To describe a gradual increase, we use the following function from \citet{Dehnen2000}:
\begin{equation}\label{timedepeq}
\psi_2 = \psi_{max}(\frac{3}{16}\xi ^5 - \frac{5}{8} \xi ^3 +\frac{15}{16} \xi +\frac{1}{2}), \, \, \, \xi \equiv 2 \frac{t}{t_{grow}} -1.
\end{equation}
Where $-1 \leq \xi \leq 1$ and $\psi_2$ varies smoothly and has null derivatives at both ends of the range. 
In order to get an idea of how slow the warping process should be, we run some test simulations with one particle at an initial circular 
orbit at 14 kpc from the Galactic centre. In Figure \ref{fig:orbit} we plot the integrated path of this particle while the potential is gradually being warped.
The warping is introduced progressively through a time  $t_{grow} \,= \, n \times P$ where P is the orbital period of a star with a circular orbit at 20 kpc in the A$\&$S potential. In the quickly warped potential, the particle acquires a sizeable amount of motion orthogonal to the instantaneous plane of the warped 
disc potential, whereas in the adiabatic case, the particle moves along this warped disc. In order to warp the potential in the adiabatic regime
we choose $t_{grow}\,=\, 6P$ which is about $3.5$ Gyr\footnote{This is not meant to be an actual time-scale in which a warping may develop in a real galactic disc, as we are not simulating an actual warping mechanism. Rather, this is the time-scale for our warping transformation meant to provide us with a reasonable set of particles within a warped potential}.\\
We use a 7-8 th order Runge-Kutta integrator with adaptive time step \citep[]{dopr}. The integration is done in three stages. First the test particles get relaxed in the A$\&$S potential, the integration time at this stage 
depends on the velocity dispersion of the tracer, the colder they are, the more time they need to reach the statistical equilibrium 
with the potential. We integrate RC stars for $10$ Gyr, A type stars for $20$ Gyr and OB stars for $30$ Gyr. Next, we adiabatically 
warp the potential for $t_{grow}\,=\,3.5$ Gyr.
 In the last part, we integrate the particles in the final state
of the warped potential for $t_2 \,=\, 2P$ to let the particles relax in this newly warped potential\footnote{Again, we do not pretend this is an actual time during which real stars 
have orbited unperturbed around the Galaxy. This is just a time to relax our initial conditions within the assumed potential.}. During the stages where the potential 
is time independent, energy conservation was better than $0.01\%$ in all cases. For the cases where we want to generate a twisted warp, 
twisting the line of nodes (see Sec. \ref{subsec:twist}) is done at this point, i.e. after finishing the integration.
 The samples we get here are the \emph{Perfect samples} to which we will refer later in this paper.

\begin{figure}
\begin{center}
\includegraphics[width=60mm]{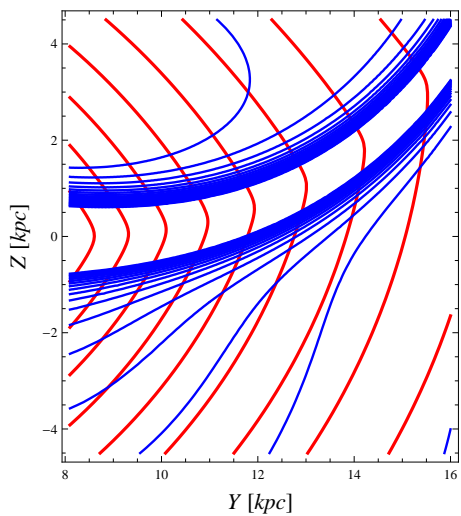}
 \caption{The equipotential and isodensity contours of the warped Miyamoto-Nagai disc potential of the A$\&$S model
respectively in red and blue. The plot is done in the Y-Z plane for X=0. It is clear that the position of the warped plane coincides
 for both, warped potential and warped density function.}
 \label{fig:equipot}                              
\end{center}
\end{figure}

A point that must be considered is that, when applying a geometrical transformation
to a potential, the corresponding density (in the sense of being proportional to the
Laplacian of the transformed potential) does not coincide with the density that is
obtained from applying the transformation to the original density function. This
may introduce a discrepancy, in the sense that the parameters of the estimated
warp (obtained using the test particle distribution) do not coincide with those that
were applied to the potential.  However, we must point out that we are only recovering
the tilt and twist angles that define the transformed planes of symmetry of the original functions
and these do coincide, as the Laplacian preserves these planes. This is shown in figure \ref{fig:equipot},
where we present the equipotentials of the warped Miyamoto-Nagai disc potential (UWF model), together
with the isodensity contours of the corresponding density function.

%-----------------------------------------------------------------------------------------------
\subsection{The Gaia ``observed samples''}
\label{subsec:obsamp}

\subsubsection{The Gaia selection function}
\label{subsec:selecfunc}
We want to study how Gaia can improve our knowledge of the warp. Then it is necessary to determine which ones of our simulated stars
 can be observed by Gaia. 

Measuring the unfiltered (white) light in the range of 350--1000 nm, Gaia yields G magnitudes. Stars brighter
 than $G=20$ 
can be observed. The Radial Velocity Spectrometer (RVS) implemented inside Gaia, provides radial velocities through 
Doppler-shift measurements. This instrument will integrate the flux of the spectrum in the range of 847--874 nm (region of the CaII triplet)
 which can be seen as measured with a photometric narrow band yielding $G_{RVS}$ magnitudes. 
These measurements are collected for all stars up to $G_{RVS}=17$.

\begin{figure*}
\begin{center}
\includegraphics[width=160mm]{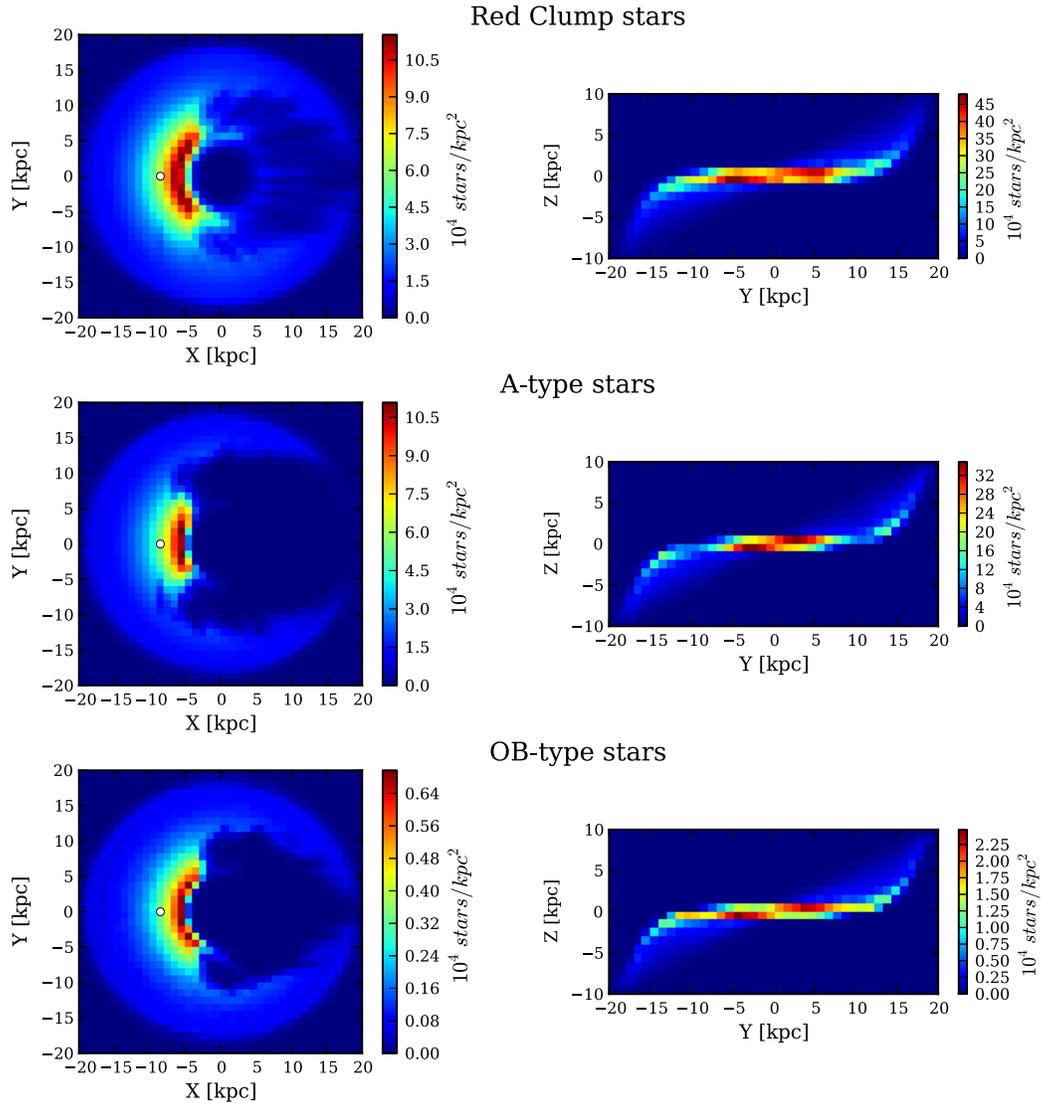}
 \caption{The distribution of stars of each warped tracers (warped with UWF model) in X-Y (left panels) and Y-Z (right panels) planes.
These are the stars observable with Gaia ( i.e. with apparent magnitudes $G<20$). The distribution of RC, A and OB stars are shown in
 the top, middle and bottom panels respectively. 
The colour scale indicates the surface density ($10^4$ stars /  $\mathrm{kpc}^{2}$). This scale is different for each tracer 
population to better illustrate the surface density. The 3D extinction map
of \protect \cite{drimmel03} is used for calculating the apparent magnitudes. Note that the Sun is located at $(x,y,z)=(-8.5,0,0)$ kpc
 as labelled with a white filled circle in the left panels. 
Therefore, in the Y-Z projection, the Sun is projected on top of the Galactic centre at Y=0. It is worth mentioning that the distances used 
in this plot are true distances.}
 \label{fig:warp_density}                              
\end{center}
\end{figure*}

For the sample whose kinematics mimics RC stars, we assign an absolute magnitude of  $M_k=-1.61$ \citep[] {alves2000} and 
an intrinsic colours of $(J-K)_o=0.55$ \citep[] {straizys09}, $(V-I)_o=1.0$ and $(V-K)_o=2.34$ \citep[] {alves2000} to each star.
 We calculate the visual absorption using the 3D extinction map of \cite{drimmel03} using rescaling factors. 
The rescaling factors are used to correct the dust column density 
of the smooth model to account for small scale structure not
described explicitly in the parametric dust distribution model. Using the extinction laws from \cite*{cardelli}, 
we calculate the apparent K magnitude and observed colour for 
every individual star.
The G magnitude is then calculated as a function of K apparent magnitude and $(J-K)$ observed colour as follows 
(J.M. Carrasco, private communication):
\begin{eqnarray}
G=\,K\,-0.286\,+4.023(J-K)\,-0.35(J-K)^2 \nonumber \\
\,+0.021(J-K)^3
\end{eqnarray}
\begin{eqnarray}
G_{RVS}=\,K\,-0.299\,+2.257(J-K)\,+0.042(J-K)^2 \nonumber \\
 \,-0.002(J-K)^3
\end{eqnarray}
For A stars, the visual absolute magnitudes are assigned using the luminosity function from \cite{murray}
 to generate stars in the range of $[A0,A9]$.
The $(V-I_C)$ colours are obtained from the absolute magnitude and colour relation presented in \cite{kenyon}.
For OB stars, the visual absolute magnitudes are obtained using the luminosity function from \cite{mottram}. 
The corresponding  $(V-I_C)$ colours are calculated using the absolute magnitude, effective temperature and colour 
relations presented in \cite{mottram} and \cite{jordi2010}. The G and $G_{RVS}$ magnitudes for 
the two latter mentioned tracers are calculated using the third order polynomial fit of $V$ apparent magnitudes and $(V-I_C)$ 
observed colours of \cite{jordi2010}. Considering stars with $G<20$, in Figure \ref{fig:warp_density} we plot the surface density of stars in X-Y and Y-Z 
cartesian projections for the three tracers. Looking at the Y-Z projection, the warped shape of the disc is clearly seen even after 
applying the Gaia selection function. From the X-Y projections it is clear that A stars are less visible to Gaia in the first
 and fourth quadrants compared to RC and OB stars since they are not as bright as OB stars, or as red as RC stars.
 As expected, the RC population has a larger scale height as seen in Y-Z projection.
\subsubsection{The Gaia error model}
\label{subsec:Gaiaerror}
\begin{figure}
\begin{center}
\includegraphics[width=90mm]{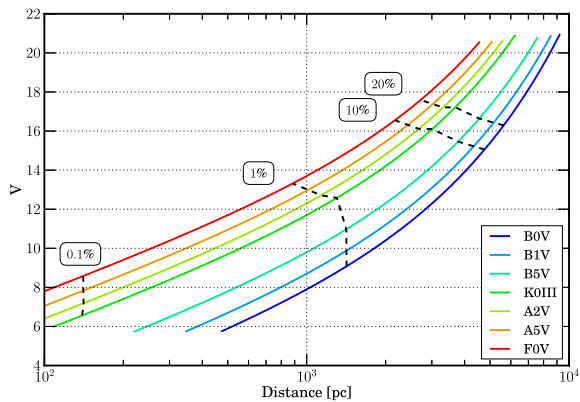}
 \caption{The mean parallax accuracy horizons for stars of different spectral type that represent our three tracer populations. 
The plot of visual apparent magnitude versus heliocentric distance is done assuming an extinction of 1 magnitude per 1 kpc.
Dashed lines represents the constant line of mean relative parallax accuracy. Note that the lines of fixed relative error
for stars brighter than $V\sim12$ are almost vertical due to the Gaia observing strategy (gates are introduced to avoid 
saturation). }
 \label{fig:par_hor}                              
\end{center}
\end{figure}

The Gaia web-page\footnote{http://www.cosmos.esa.int/web/gaia/science-performance} provides science performance estimates 
and models for errors in astrometric, photometric and spectroscopic data. The end-of-mission parallax errors depend
 on G magnitudes and $(V-I_C)$ colours. Figure \ref{fig:par_hor} shows the mean parallax accuracy horizons for stars with different
 spectral types that represent our three tracers.
 We also take into account the variation of errors as a function of ecliptic coordinates, due to the variation of number of transits at the end of the mission.
 The Galactic coordinates and proper motion errors are described as a function of the error in parallax. The end-of-mission radial 
 velocity errors depend on V magnitudes and the spectral type of the stars. The redder they are, the smaller error in their radial 
velocity measurements. Applying the 3D extinction map, Gaia observational constrains and these error models to our perfect sample, 
we generate the Gaia ``observed catalogues'' for the three tracers.
\subsubsection{Characteristics of the observed samples with and without velocity information}
\begin{figure*}
\begin{center}
\includegraphics[width=180mm]{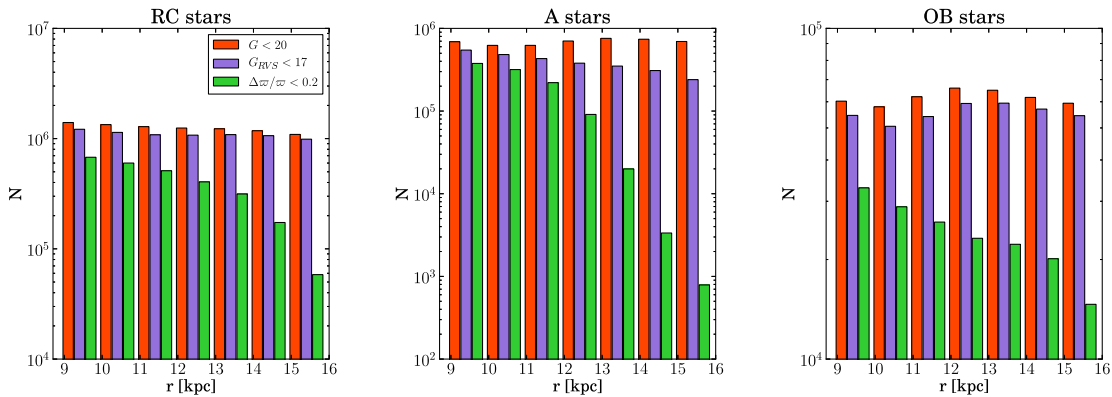}
 \caption{Histograms of the number of stars in Galactocentric radius bins of 1 kpc. The samples with $G<20$, $G_{RVS}<17$ and $\Delta \varpi / \varpi <0.2$ are shown respectively in red, purple and green. 
The histograms are plotted for RC stars (left panel), A stars (middle) and OB stars (right). }
 \label{fig:tracer_hist}                              
\end{center}
\end{figure*}

It is essential for our study to know how many stars will be observed by Gaia as a function of Galactocentric radius for each tracer population.
 We also have to know for how many of them Gaia provides radial velocity information and also how many with good parallax measurements.
Figure \ref{fig:tracer_hist} shows the histograms in logarithmic scale of the number of stars in (spherical) Galactocentric radius bins of 1 kpc, 
starting from 9 kpc up to 16 kpc for the three tracers. 
 It is worth noting that for the samples
with error in parallax less than 20$\%$ ($\Delta \varpi / \varpi <0.2$) the number of A stars drops down by three orders of magnitude
at large radii while this reduction is only one order of magnitude and less than one for RC and OB stars respectively.
\label{subsec:characsamples}
%-----------------------------------------------------------------------------------------------
\section{Warp detection and characterization}
\label{sec:warpdetect}
\subsection{The mGC3 method}
\label{subsec:mgc3_method}
\begin{figure*}
\includegraphics[width=90mm]{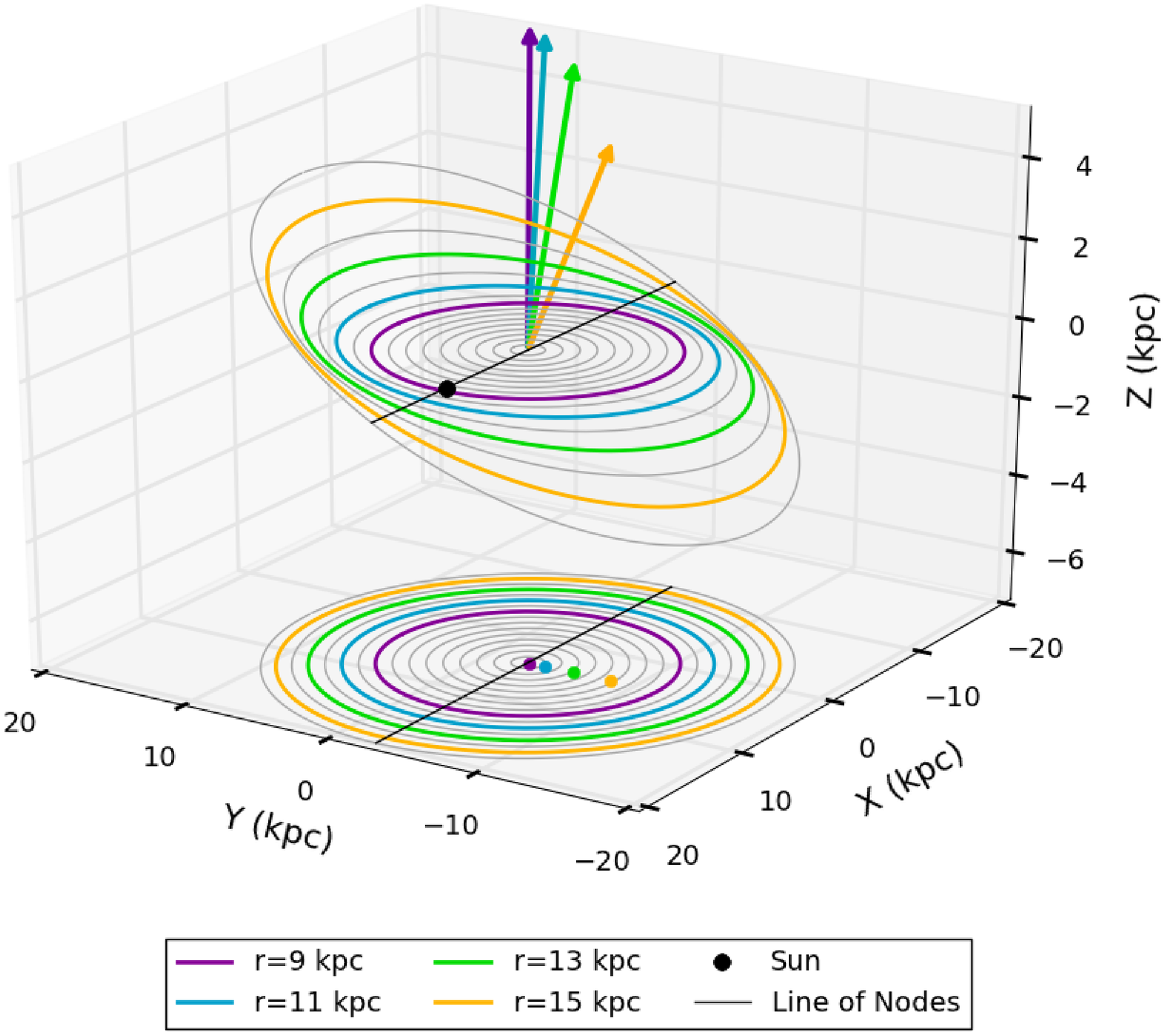}
\includegraphics[width=70mm]{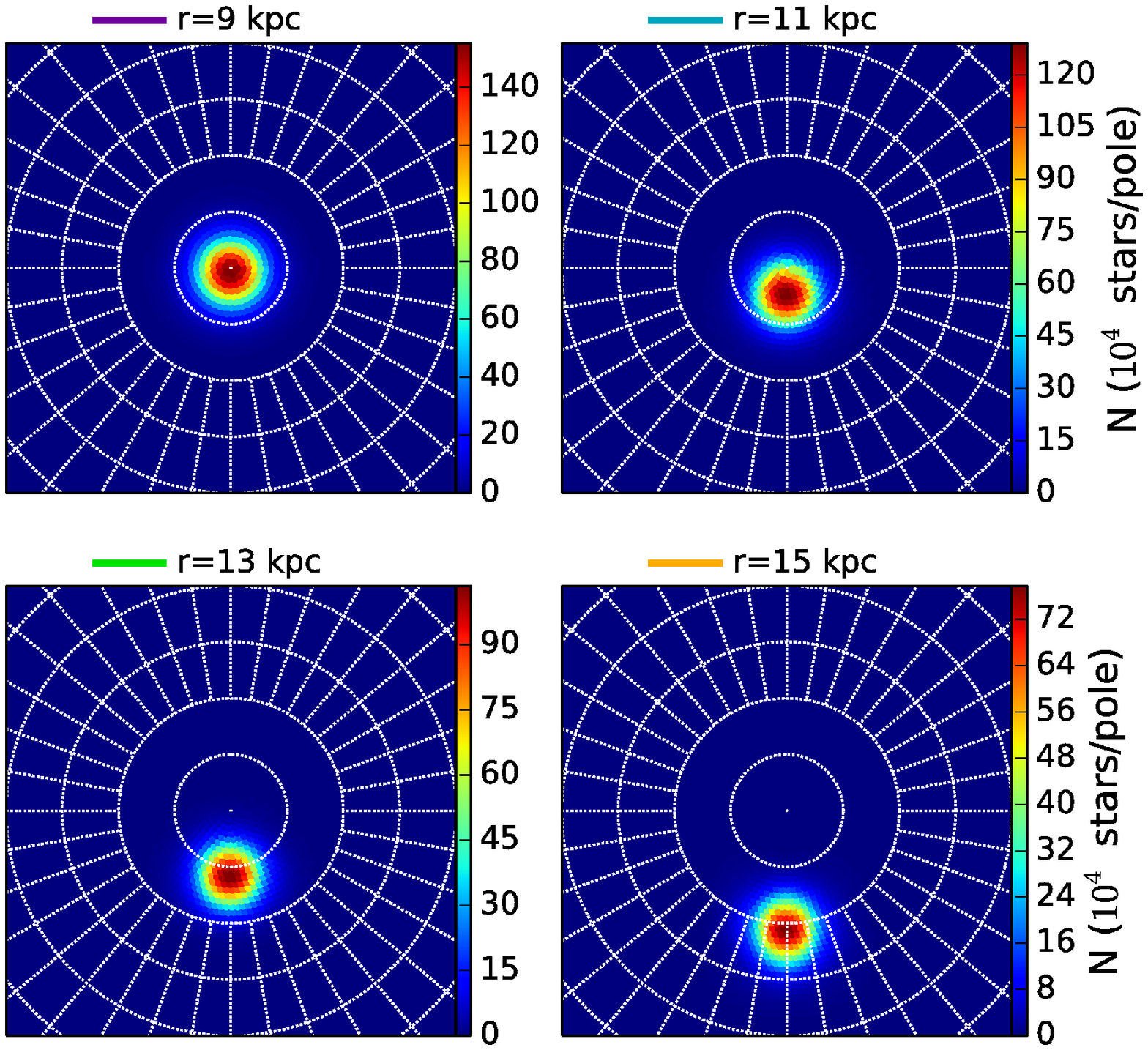}
 \caption{\emph{Left:} Schematic plot of a warped disc (UWF model). 
The solid black line and black dot indicate respectively the line of nodes and the Sun's position. Concentric circles shown in gray 
increase in Galactocentric radius by 1 kpc, with colours showing four selected rings at 9 (purple), 11 (blue), 13 (green) and 15 kpc 
(orange). The arrows represent the normal vectors corresponding to each of these rings, in the same colour. The projection in the X-Y 
plane shows the tips of the normal vectors, aligned along the direction perpendicular to the line of nodes, which in this case is a 
straight line. \emph{Right:} The $2\times2$ mosaic shows the pole count maps which correspond to stars in spherical shells with the 
same radii as the coloured rings indicated in the left panel. The simulated A stars are used to generate these plots. 
The maps are shown in an a north-polar azimuthal equidistant projection,
 showing the north pole at the centre, $\varphi=90\degr-270\degr$ in the vertical direction at the centre of the plot, the concentric
 circles have a separation of $5\degr$,  and meridians are drawn at $10\degr$ intervals in longitude. The colour scale shows the number of stars associated to each pole according to the mGC3 criteria 
of Equation \ref{eqn:mgc3_criteria}. Each of the pole vectors indicated by the arrows in the left panel correspond to the pole with maximum 
counts in the pole count map for the respective radius. The sequence of plots clearly shows how the position of the pole with the maximum
 star counts shifts in latitude as the radius $r$ increases, as do the arrows shown in the left panel, as a consequence of the increase 
in the tilt angle $\psi(r)$. The azimuthal angle $\varphi$ of the maximum counts pole (and the corresponding arrows) remains constant as
 expected for a straight line of nodes.}
 \label{fig:warp_scheme_pcm_seq_notwist}
\end{figure*}

The modified Great Circle Cell Counts method (mGC3) was introduced by \citet{mateu11} as a technique for the detection of tidal streams
 in the Galactic Halo, based on the 
original GC3 method proposed by \citet*{johnston1996} for the same purpose. The mGC3 method is based on the fact that,
in a spherical potential, the tidal stream produced by the disruption of a satellite in the Galactic Halo
will conserve its total angular momentum and its orbit will be confined to a plane, which will be an exactly constant plane if the
 potential is perfectly spherical, or will precess if it is axisymmetric.
Therefore, as seen from the Galactic centre, the stars in the stream are confined to a great circle band, the projection of the orbital
 plane. This means that \textit{both} the Galactocentric position and velocity vectors of stream stars are perpendicular, within a
 certain tolerance, to the normal vector or \emph{pole} $\mathbf{\hat{L}}$ which defines this particular great circle. 

The mGC3 method thus consist in producing a \textit{pole count map}\footnote{The Python package PyMGC3 containing code to run the mGC3/nGC3/GC3 family of methods is publicly available at the github repository https://github.com/cmateu/PyMGC3}, i.e. a map of the number of stars associated to each possible 
pole (and  therefore great circle cell) in a grid in spherical coordinate angles (directions), computed as the number of stars which,
 for each pole, fulfil the following criteria

\begin{equation}\label{eqn:rvcriteria}
|\mathbf{\hat{L}}_i \cdot \mathbf{\hat{r}}_\mathrm{gal}| \leq \delta_r \quad \mathrm{and} \quad  |\mathbf{\hat{L}}_i \cdot \mathbf{\hat{v}}_\mathrm{gal}| \leq \delta_v 
\end{equation}

where $\mathbf{\hat{r}}_\mathrm{gal}$ and $\mathbf{\hat{v}}_\mathrm{gal}$ are unit vectors in the direction of the star's 
(Galactocentric) position and velocity vectors respectively, $\delta_r$ and $\delta_v$ are the tolerances and $\mathbf{\hat{L}}_i$ is 
the $i$-th vector on the grid of all poles considered. The modification mGC3 introduces to the original GC3 
method is the use of the velocity criterion in Equation \ref{eqn:rvcriteria}, which as shown in \citet{mateu11},
 increases the efficiency of the method by reducing substantially the background contamination. Also,
in order to avoid the distance bias introduced by the reciprocal of the parallax \citep{brown05}, 
\citet{mateu11} use the criteria of Equation \ref{eqn:rvcriteria} expressed in the following equivalent manner

\begin{equation}\label{eqn:mgc3_criteria}
|\mathbf{\hat{L}}_i \cdot \mathbf{r'}_\mathrm{gal}| \leq \Vert \mathbf{r'}_\mathrm{gal} \Vert\delta_r \quad \mathrm{and} \quad  |\mathbf{\hat{L}}_i \cdot \mathbf{v'}_\mathrm{gal}| \leq  \Vert \mathbf{v'}_\mathrm{gal} \Vert \delta_v 
\end{equation}

where $ \mathbf{r'}_\mathrm{gal}$ and $ \mathbf{v'}_\mathrm{gal}$ are simply the position and velocity vectors  $\mathbf{r}_\mathrm{gal}$ and $ \mathbf{v}_\mathrm{gal}$, multiplied by the parallax, which in terms of the heliocentric observable quantities $(l,b,\varpi,v_r,\mu_l,\mu_b)$ are given by

\begin{equation}\label{eqn:rgal_vgal_prime}
\begin{array}{l}
 \mathbf{r'}_\mathrm{gal} = \varpi\mathbf{r}_\odot + A_p\big((\cos{l}\cos{b}) \mathbf{\hat{x}}+(\sin{l}\cos{b}) \mathbf{\hat{y}}+(\sin{b}) \mathbf{\hat{z}}\big)  \\
\mathbf{v'}_\mathrm{gal} =  \varpi\mathbf{v}_\odot + \varpi v_r \mathbf{\hat{r}} + (A_v \mu_l \cos{b}) \mathbf{\hat{l}}+ (A_v\mu_b) \mathbf{\hat{b}}
\end{array} 
\end{equation}

where $A_p = 10^3$ $\mathrm{mas\, \,pc}$, $A_v = 4.74047$ $\mathrm{yr \,kms}^{-1}$, $\{\mathbf{\hat{x}},\mathbf{\hat{y}},\mathbf{\hat{z}}\}$ are cartesian unit vectors
 and $\{\mathbf{\hat{r}},\mathbf{\hat{l}},\mathbf{\hat{b}}\}$ are the unit vectors in heliocentric Galactic coordinates 
\citep[for full details see][]{mateu11}.

The mGC3 method is ideally suited to study and characterize a warp with a fixed tilt angle for each Galactocentric ring, similar to 
our untwisted warp models.
For a flat disc, pole count maps made for stars in bins with increasing (Galactocentric) distance $r$ will show a maximum located
 always at the North Galactic Pole. For a tilted ring model of a warped disc (see Sec. \ref{subsec:warp}), each of the rings has a 
tilt angle $\psi(r)$ which will yield maximum counts at a pole  located at a latitude $\theta=\pi/2-\psi(r)$. 

Figure \ref{fig:warp_scheme_pcm_seq_notwist} illustrates the pole count maps obtained when applying the mGC3 method to the sample
of A stars, warped with the UWF.
 The left panel shows a 3D schematic plot of the warp model. Concentric rings in
 gray show the mid-plane of the warped disc, with radii increasing in steps of 1 kpc. 
The line of nodes is shown as a black solid line and the position of the Sun indicated with a filled black dot.
Four particular rings (at 9, 11, 13 and 15 kpc) and their respective pole vectors are emphasized in colours (see figure caption and 
legend). The tips of the pole vectors are indicated as filled circles in the X-Y plane projection, with the same colours. The plot 
shows how the pole vectors deviate from the Z-axis with an angle equal to the tilt angle of the respective ring, which increases 
with Galactocentric $r$. Since there is no twisting, the pole vectors have a constant azimuthal angle $\varphi$ and their tips are 
distributed along a straight line in the X-Y plane projection. The right part of the plot shows a $2\times2$ mosaic, with the pole 
count maps  corresponding to each of the four coloured rings shown on the 3D plot (left). For generating these pole count maps we use the 
full sample of A stars, warped with UWF.  The top left panel shows the map for the 
innermost ring, at a radius of 9 kpc, which has a very small tilt angle ($\psi=0.2\degr$) and the maximum pole counts located almost 
exactly at $\theta=90\degr$, at the very centre of the map. As the radius increases, the maximum counts signature moves towards lower 
latitudes as $\theta=\pi/2-\psi(r)$, exhibiting the dependence of the tilt angle with distance. Note also the maximum pole counts
 displace only in latitude, as expected for a warp with no twisting, remaining at a constant azimuthal angle which defines the direction
 perpendicular to the line of nodes.

\begin{figure*}
\includegraphics[width=90mm]{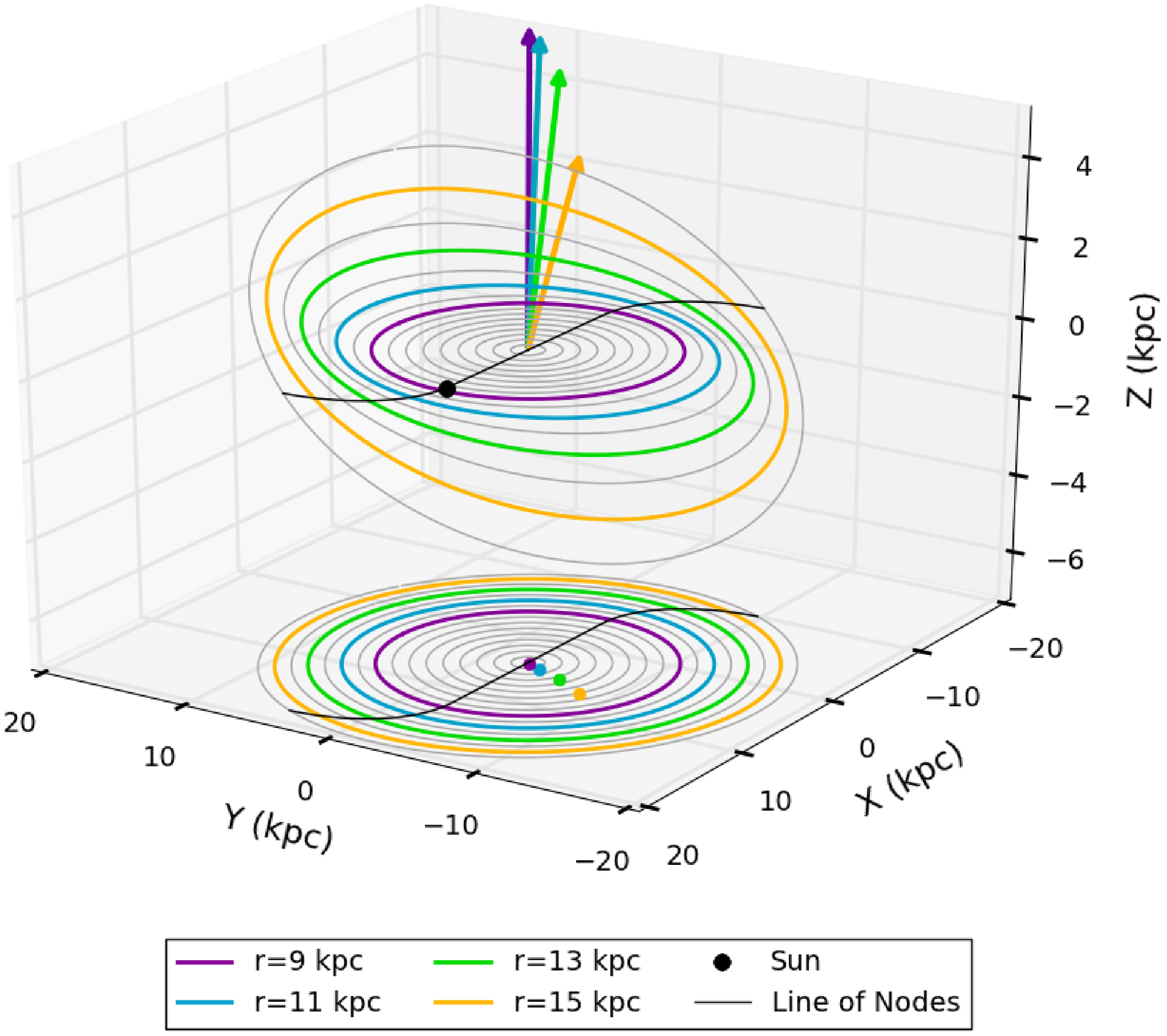}
\includegraphics[width=70mm]{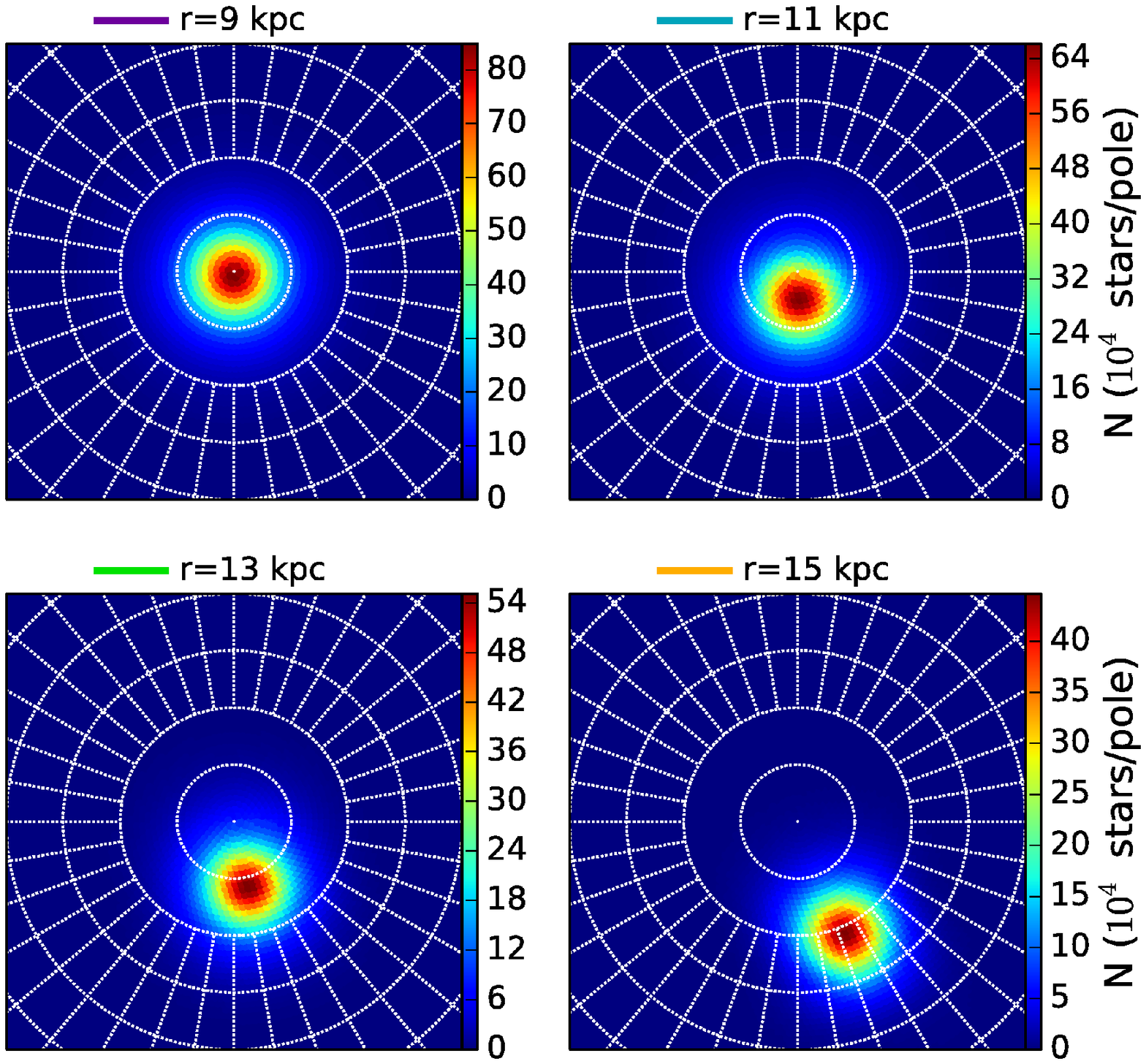}
 \caption{Same as Figure \ref{fig:warp_scheme_pcm_seq_notwist} for a warp model with a twist. \emph{Left:} Schematic plot 
of a warped disc of TW2.
 The projection in the X-Y 
plane shows the tips of the normal vectors, which now deviate from the Y=0 axis with an angle that increases proportionally with radius
 due to the twisting of the line of nodes. \emph{Right:} 
 The sequence of pole count maps shows how, in 
addition to the shift in latitude caused by the tilting of the rings, the position of the pole with the maximum star counts now also 
shifts in the azimuthal direction with $\varphi$ increasing as $r$ does, illustrating the twisting of the line of nodes as a function
 of radius depicted in the left panel.}
 \label{fig:warp_scheme_pcm_seq_twist60}
\end{figure*}

A warped model \textit{with} twisting is illustrated in Figure \ref{fig:warp_scheme_pcm_seq_twist60}, with the same layout and colour 
scheme as in Figure \ref{fig:warp_scheme_pcm_seq_notwist}. The warp model shown is the TW2 (see Sec. \ref{subsec:twist}). 
The left plot shows how the line of nodes is twisted for radii larger than 8 kpc, as seen also in the X-Y projection where the tips of 
the pole vectors are shown to 
deviate from the straight line these followed in Figure \ref{fig:warp_scheme_pcm_seq_notwist}. The array of pole count maps illustrates
 how in this case, in addition to the displacement in the latitudinal direction, now the maximum counts
 pole also shifts in the azimuthal direction with an angle that increases with distance as the twist angle does. Here the RC stars
sample that is warped with TW2 is used for generating the pole count maps.

It is worth emphasizing that mGC3 maps provide a means for empirically measuring the tilt and twist angles
of the warp as a function of distance, in a completely non-parametric way, without making any assumptions on the functional form of
 this dependence. The only assumption that is implicitly being made is that the warp is symmetric, in the sense of the tilt angle being
 the same on either side of the line of nodes, which will produce a single peak in the pole count maps. Lopsided warps may produce very
 different signatures on mGC3 pole count maps, depending on the actual shape of the warp. We will discuss the behaviour of mGC3 maps with
 a lopsided warp model in the next paper of this series. 

\subsection{The new nGC3 method}
\label{subsec:gc3_methods}
Since the use of mGC3 requires all six-dimensional phase-space information, it is also worth while exploring the performance of the method when  introducing a couple of variations, when less information is available. The largest restriction when using all positional and kinematical information comes from the magnitude limit set for the measurement of radial velocities by Gaia, which restricts the sample to $G_{RVS}<17$  (see Sec. \ref{subsec:selecfunc}). Therefore, we also explore the performance of mGC3 omitting the radial velocity term, $\varpi v_r \mathbf{\hat{r}}$, in  Equation \ref{eqn:rgal_vgal_prime} and using only proper motion information. In this way we trade less kinematical information
 for a larger sample. In the following analyses we will refer to this 
new variation as \textit{nGC3} (no-radial velocity mGC3). Additionally we produce pole count maps using only the positional 
criterion in Equation \ref{eqn:mgc3_criteria}, i.e. using the GC3 method\footnote{The original GC3 method as devised 
by \cite{johnston1996} uses heliocentric coordinates $(l,b)$ rather than Galactocentric. Here we use a Galactocentric GC3 method}. 

It is clear that the lack of radial velocity information will introduce limitations.
The use of proper motions in nGC3 helps in reducing the contamination in comparison to GC3, which uses only positional information, though lacking the information provided by the radial velocity necessarily implies that there will still be some contamination left over.
The contamination will have a larger effect in different directions, depending on whether or not the radial velocity component of stars in the feature we're interested in (in this case the disc) has a large contribution to the full velocity vector or not. For instance, for disc stars in the direction towards the Galactic centre or anticentre ($l=0\degr,180\degr$), the radial velocity component is negligible and so the full velocity vector is well approximated by its tangent velocity. The position and velocity vectors of stars in these directions give a good handle on the definition of a preferential plane of motion.
On the other hand, in the perpendicular direction ($l=90\degr,270\degr$), the projection of the velocity vector of disc stars in the radial direction is not negligible at all, which means stars in these directions are more prone to contamination and thus the identification of a preferential plane is more uncertain.
In spite of this, we deem this effect to be negligible since our interest lies on disc stars which by far dominate star counts in all directions. Contamination from Halo stars is expected to be at most $\sim1\%$  \citep[]{carney}, based on solar neighbourhood star counts, but it would even vanish when using a tracer such as OB stars, present only in very young populations; or even MS A-type stars, more massive than the Halo F turn-off. 

\begin{figure*}
\includegraphics[width=160mm]{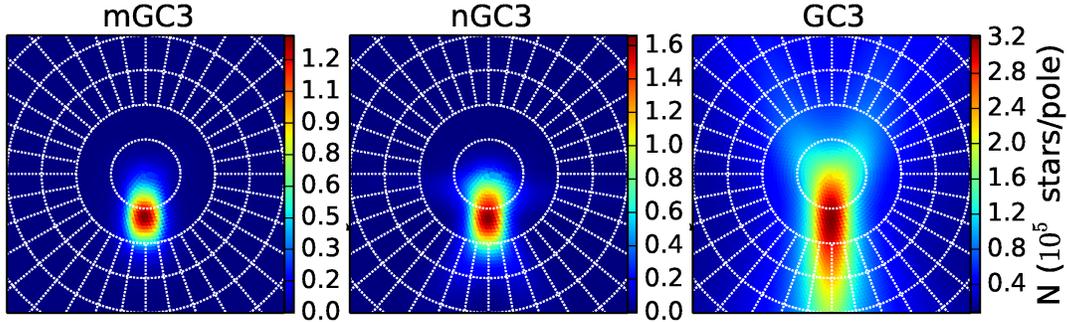}
 \caption{Pole count maps produced with the mGC3 (\emph{left}), nGC3 (\emph{centre}) and GC3 (\emph{right}) methods, for the 13 kpc
 ring shown in the warp model of Figure \ref{fig:warp_scheme_pcm_seq_notwist}. The colour scale indicates the number of stars associated
 to each pole. Note that the warped A stars sample with UWF, is used for generating these pole count maps. }
 \label{fig:mnogc3_maps}
\end{figure*}

We expect the use of kinematical information to yield a cleaner peak signature in mGC3 and nGC3 pole count maps. This is illustrated 
in Figure \ref{fig:mnogc3_maps}, where we show mGC3, nGC3 and GC3 pole count maps, from left to right, for the 13 kpc ring in the warp model illustrated in Figure \ref{fig:warp_scheme_pcm_seq_notwist} using the A stars sample.
 The colour bar shows the number of stars associated with each pole for
 each of the maps. As expected, the number of star counts is much larger for the GC3 and nGC3 methods, which allow the use of all stars
 up to the Gaia  limiting magnitude $G=20$. It is clear that nGC3 and mGC3 maps provide a much narrower and well-defined peak 
with nearly zero background counts, providing a clear example of the great benefit of having proper motions; whereas the peak in the GC3 pole count map is more extended, in particular in the latitude direction, and is embedded in a higher and noisier background. In the following analysis we will evaluate the performance of these three methods 
in the recovery of the tilt and twist angles for different warp models and tracer populations.

\subsection{The peak finder procedure}
\label{subsec:peakfinder}

In order to identify the position $(\varphi_o,\theta_o)$ of the peak in a pole count map, we have used a Bayesian framework which
 provides a straightforward way to compute also the associated uncertainties.

First, we switch to a cartesian north-polar azimuthal equidistant (NPAE) projection of the pole count maps (such as those of Figures 
\ref{fig:warp_scheme_pcm_seq_notwist}-\ref{fig:mnogc3_maps}), rather than using an Aitoff projection of the spherical coordinates, in
 which the peak finding analysis is more difficult due to the curvature inherent to the coordinate system. 
Since we are only interested in finding the position of the peak and not in modelling its entire shape accurately, we assume it can
 be described by a simple two-dimensional Gaussian in the Cartesian projection, which we express as

\begin{equation}
N_M(x,y)= A e^{-\frac{(x-x_o)^2}{2\sigma_x^2}}e^{-\frac{(y-y_o)^2}{2\sigma_y^2}}
\end{equation}

where we have assumed independence in $x$ and $y$ by omitting the crossed terms $\sigma_{xy}$ of the covariance matrix. Also, we
 restrict the sample to grid points with pole counts
higher than 60$\%$ of the maximum counts, since we only intend to use the Gaussian model
 in the area right around the peak. For the observed pole counts $N_i$ we assign typical Poisson counting errors $\sigma_i=\sqrt{N_i}$. We assume these errors to follow a 
Gaussian distribution \footnote{This is a reasonable assumption since typically $N$ is very large ($>10^4$).},
and thus we express the logarithm of our likelihood function $L=p(\{N\}|x_o,y_o,\sigma_x,\sigma_y,A)$ as

\begin{equation}\label{eqn:likelihood}
 \ln{L} = \sum_{i=1}^{n} -\frac{(N(x_i,y_i)-N_M(x_i,y_i))^2}{2\sigma_i^2}
\end{equation}

which gives the probability of having observed a set $\{N\}$ of pole count measurements, for a given set of model parameters 
$\{x_o,y_o,\sigma_x,\sigma_y,A\}$. 

We assume uniform prior probability distributions for all our model parameters, in the following allowed ranges: $A$ between 
the minimum and ten percent plus the maximum observed counts; $x$ and $y$, between the minimum and maximum values given by the 
Cartesian pole grid; $\sigma_x$ and $\sigma_y$, between zero and half the range spanned by the Cartesian pole grid. Our posterior 
probability $p(x_o,y_o,\sigma_x,\sigma_y,A|\{N\})$ is then simply proportional to the likelihood in Equation \ref{eqn:likelihood}.

Samples from the posterior distribution were obtained
 by using the Markov Chain Monte Carlo (MCMC) sampler \emph{emcee} from \citet{foreman-mackey2013}, which provides a Python
 implementation of an affine-invariant
MCMC sampler \citep{goodman2010}. The \textit{emcee} sampler has the advantage of providing an MCMC algorithm with very few 
free parameters \citep[see][for full details]{foreman-mackey2013}: the number of `walkers' (the number of simultaneous chains 
to be used), the number of burn-in steps and the total number of final chain steps. The sampler parameters were set to $300$ 
walkers with $150$ and $200$ burn-in and total steps respectively, which resulted in acceptance fractions in the range $\sim0.25-0.4$ 
and auto-correlation times $\sim15$ times smaller than the total duration of the chain, well within the ranges suggested 
by \citet{foreman-mackey2013}.

Finally, the MCMC $(x,y)$ values are transformed back to spherical coordinates $(\varphi,\theta)$. For this distribution 
of $(\varphi,\theta)$, which corresponds to a sampling of the marginalized posterior probability $p(\varphi_o,\theta_o|\{N\})$, 
we compute the median, $15.8$th and $84.2$th percentiles which we take respectively as the best estimate  and the asymmetric $1\sigma$ 
confidence intervals for $(\varphi_o,\theta_o)$ \citep{hogg2010}.

In the following section we describe the results obtained from applying our peak finding algorithm to the pole count maps obtained 
using the different GC3 methods. For generating the pole count maps in all of the cases, we used a tolerance of $\delta_r= \delta_v=2\degr$ and a pole
 grid spacing of $0.5\degr$. We checked that using smaller values for the tolerance ($\delta_r= \delta_v=1\degr \, , \, 0.5\degr$) and the pole grid spacing ($0.25\degr$), does not change the results significantly.
%====================================================

\section{Results}
\label{sec:results}
The methods proposed in Sec. \ref{sec:warpdetect} have been applied to three type of simulated samples: 1) \emph{The Perfect samples},
 which contain every single generated star in our simulated and relaxed warp model, as described in Sec. \ref{subsec:relax}, 
that is without any observational 
constrain; 2) \emph{The Magnitude Limited samples}, including the effect of observational errors and interstellar extinction. Considering all stars 
 up to the Gaia limiting magnitude of $G=20$ for the GC3 and nGC3 methods, which only require positional information and 
proper motions respectively, and up to magnitude $G_{RVS}=17$ for mGC3, which requires the use of radial velocity information
 (see Sec. \ref{subsec:gc3_methods}); and 3) \emph{The Clean 
samples}, including only stars with relative error in parallax smaller than 20$\%$ in addition to the 
previous observational constrains. We apply the GC3, mGC3 and nGC3 methods to the RC, A and OB star samples.
  We split the samples in Galactocentric radial bins
 from 9 to 16 kpc, with a width of 1 kpc, and compute the position of the peak in the resulting pole counts maps using the procedure
 described in Sec. \ref{subsec:peakfinder}. The values obtained for the tilt and twist angle are compared to the model predictions,
 discussing, in all cases, the
 accuracy we can reach and the systematic trends present in the analysis. In Sec. \ref{subsec:Resuntwist} we show the results for the samples warped with
 the UWF model, that we consider our \textit{fiducial model}, together with the results using the UWH model. Results for the twisted warp models (TW1 and TW2) that 
applied only to the OB Clean sample are presented in Sec. \ref{subsec:ResTwistSample}.

\subsection{Results for the untwisted warp model}
\label{subsec:Resuntwist}
\subsubsection{The Perfect samples}
\label{subsec:ResGodSamples}
\begin{figure*}
\includegraphics[width=150mm]{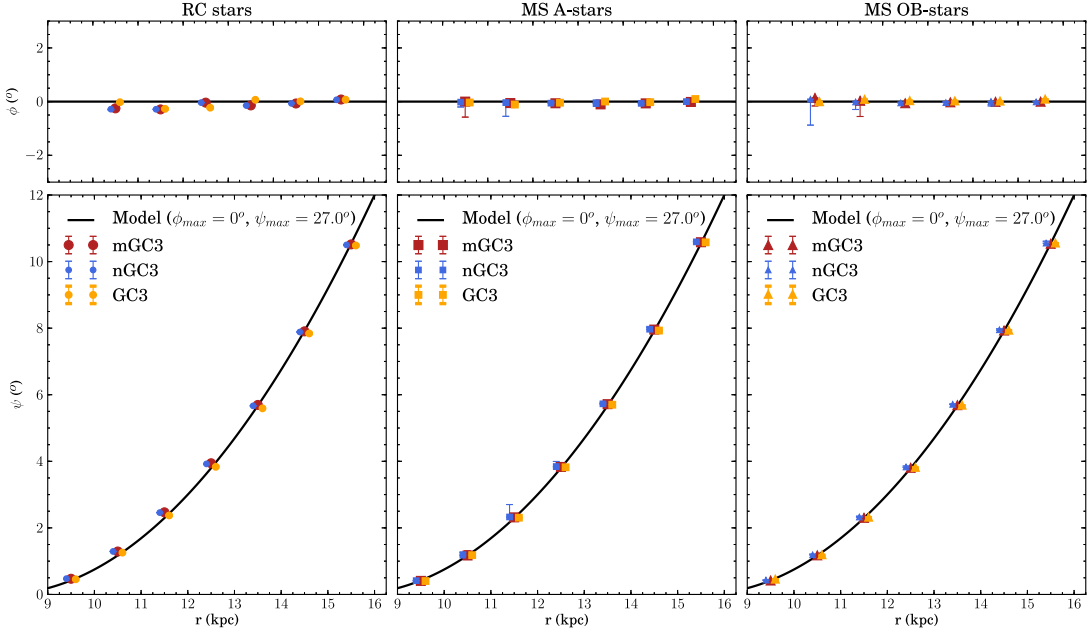}
 \caption{Tilt $\psi$ and twist $\phi$ angles versus Galactocentric (spherical) radius $r$ for the Perfect sample (i.e. without errors or selection function). 
The black solid line shows the warp model values and coloured points indicate the results obtained from mGC3 (\textit{red}),
 nGC3 (\textit{blue}) and GC3 (\textit{yellow}) pole count maps. In this plot, the points corresponding to nGC3 and GC3 have been 
shifted slightly in the horizontal direction to keep them from fully overlapping. Error bars are plotted, however for most cases, 
these are smaller than the plotting symbols. In the top panels, the point corresponding to the nearest 
bin (centred at $r=9.5$ kpc) has been omitted since for such a small expected tilt angle, the maximum counts signature is expected 
to lie almost exactly on the pole ($\theta=90\degr$), where the azimuth (twist angle) is meaningless.}
 \label{fig:tilt_vs_r_godsample}
\end{figure*}

The bottom and top panels in Figure \ref{fig:tilt_vs_r_godsample} show respectively the tilt $\psi$ and twist $\phi$ angles as a function of
 Galactocentric (spherical) radius $r$, for the Perfect sample of RC (\emph{left}), A (\emph{centre}) and OB stars (\emph{right}).
 The results are shown in comparison with the model prediction (solid black line). 
The filled red, blue and yellow points show the results obtained for the position of 
the peak obtained respectively from mGC3, nGC3 and GC3 pole count maps.  As the plots show,
 the recovery of both the tilt and twist angles for these Perfect samples is flawless, with all three methods.
This results were to be expected since these samples are error-free and are not affected by a selection function. Nevertheless 
this is not a trivial test, since we are comparing the recovered distribution of relaxed particles with the tilt and twist angles
 used to warp the potential. These results verify that the resulting distribution of relaxed particles follows a warp with the same
 $\psi(r)$ and $\phi(r)$, as those used to warp the potential.
%-----------------------------------------------------------------------------------------------
\subsubsection{The Magnitude Limited samples}
\label{subsec:ResObsSamples}

\begin{figure*}
\includegraphics[width=150mm]{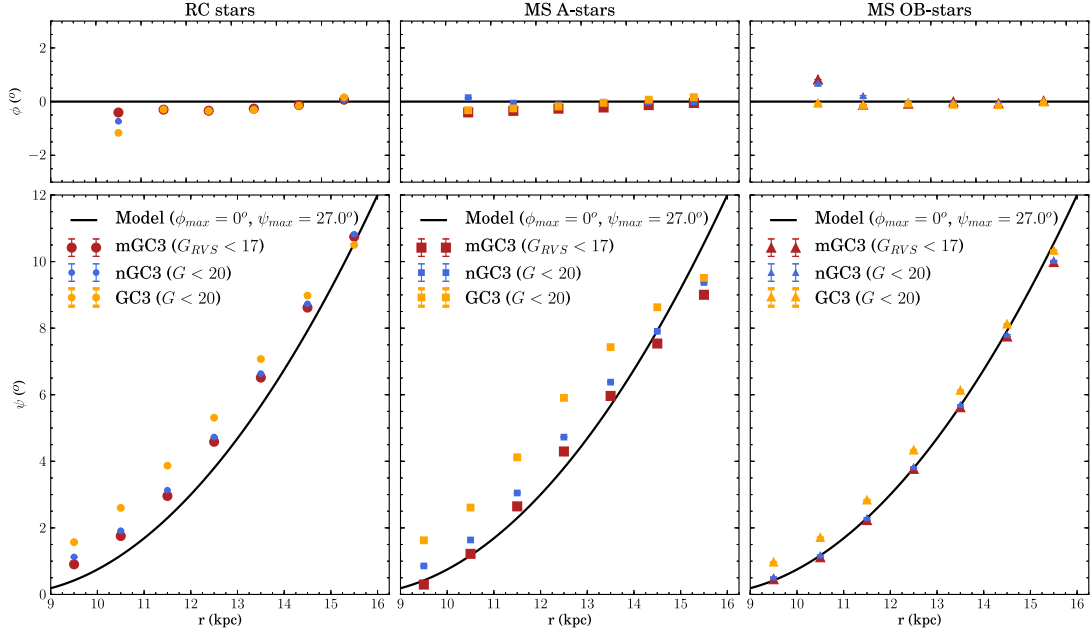}
 \caption{Tilt $\psi$ and twist $\phi$ angles versus Galactocentric (spherical) radius $r$ for the Magnitude Limited samples ($G<20$ for GC3 and nGC3, and $G_{RVS}<17$ 
for mGC3, see Sec. \ref{subsec:gc3_methods}).}
 \label{fig:tilt_vs_r_mlimsample}
\end{figure*}

We now apply the same procedure to the Magnitude Limited samples that
 includes the effects of the Gaia selection function, errors and interstellar extinction (see Sections  \ref{subsec:selecfunc}-\ref{subsec:Gaiaerror}). 
 Figure \ref{fig:tilt_vs_r_mlimsample} top panels, show that the null twist angle is recovered for all tracers,
 using all three methods. We observe very small deviations of typically less than $\sim1\degr$, except for the $10.5$ kpc
 bin for which the difference with the model is slightly higher (smaller than $\sim2 \degr$), yet less significant as we are close to 
the pole. 
The recovery of the tilt angle (Figure \ref{fig:tilt_vs_r_mlimsample} bottom panels) shows a more complex behaviour which, as expected, 
depends both on the tracer and the method selected. Two 
important factors come into play producing the observed behaviour: the 
effect of sample biases introduced by parallax errors when the samples are binned in Galactocentric radii and the intrinsic velocity dispersion of the tracer population.

The effect of the intrinsic velocity dispersion is readily observed when comparing results from different tracer populations.  
Figure \ref{fig:tilt_vs_r_mlimsample} shows that the recovery of the tilt angle is best for OB stars and less good for A and RC stars. 
This behaviour is natural since OB stars, although few, are a very kinematically cold sample and therefore, have a much smaller 
scale height, so smaller dispersion around the mid-plane of the warped disc. The A and RC star samples are more numerous, but have 
increasingly higher velocity dispersions and are, therefore, scattered farther from the mid-plane of the warped disc. It is the
 combination of this different scale height of each population with the errors in distances in the line of sight direction that 
produce the observed biases.

\begin{figure*}
\includegraphics[width=55mm]{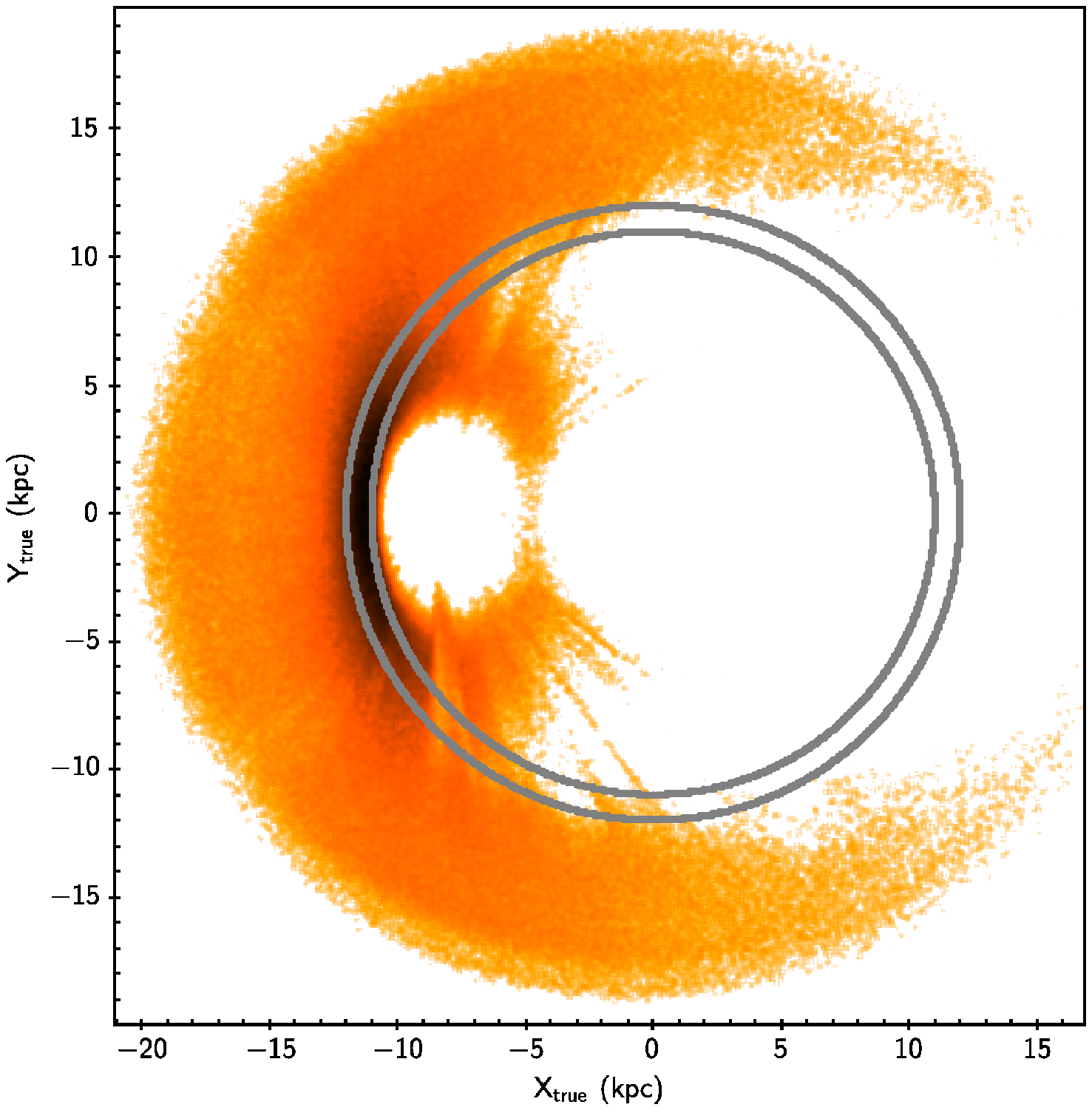}
\includegraphics[width=55mm]{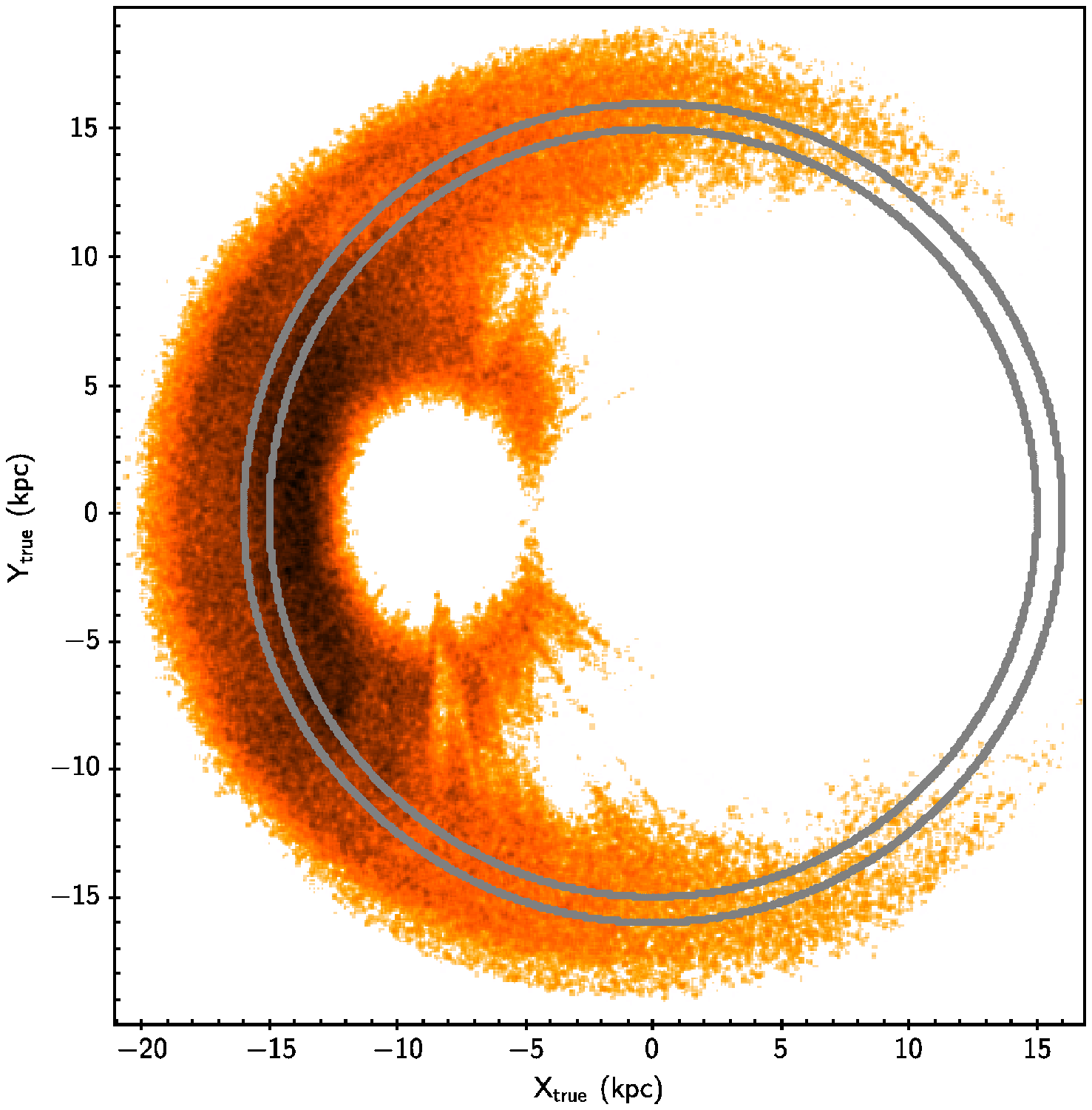}
 \caption{Plot of the \emph{true} spatial distribution in the X-Y plane, for the Magnitude Limited sample of A stars
 in the \emph{observed}  (Galactocentric) distance ranges $11<r_{obs}<12$ kpc (\emph{left}) and $15<r_{obs}<16$ kpc
 (\emph{right}). The solid gray lines indicate the higher and lower limits of the observed distance range. The Sun is 
located at (-8.5 kpc,0.) and the Galactic Center is in the origin. The colour scale is proportional to the logarithm of
 the number density, with dark colours indicating higher densities and light-orange shades indicating lower densities.}
 \label{fig:lk_bias}
\end{figure*}

To deeply understand these effects, we will now concentrate on a single tracer. Let us focus on A stars, which being the least
 luminous on average, have higher parallax errors (see Figure \ref{fig:par_hor} ). We can see that all three methods follow a common 
trend in which the tilt angle
 is overestimated up to a certain distance ( $r\sim14$ kpc for A stars ) and then it is underestimated for larger distances. This trend 
is caused by two different sample biases, which act in opposite ways and dominate at different distance ranges.
 Figure \ref{fig:lk_bias} illustrates this for A stars with \emph{observed} Galactocentric distances in the ranges $11<r_{obs}<12$ kpc (left panel)
 and $15<r_{obs}<16$ kpc (right panel). The plots show the \emph{true} spatial distribution of these stars in the X-Y plane, i.e. as
 seen pole-on from the North Galactic Pole, with a colour scale proportional to the logarithm of the number density. Note that, due to the 
small errors at short heliocentric distances, we do not have any stars with true distances close to the Sun reaching the mentioned 
observed rings. In the left panel
 it is clearly seen that the majority of the contamination comes from outside the selected $r_{obs}$, i.e. from \emph{larger distances }
with \emph{higher} tilt angles, thus biasing the mean observed tilt angle towards \emph{higher} values. This is a well known
bias caused by the combination of two effects. On one hand, the decrease of radial surface density as a function of Galactocentric distances
 causes having more stars in the inner than in the outer rings. On the other hand, moving to larger Galactocentric rings, larger volumes 
are covered. It is the combination of these two effects that makes the 
number of stars outside our distance bin much larger than the number of stars inside. These effects make it more likely for contaminants at larger 
distances to be scattered into the $r_{obs}$ bin (See red histogram in Figure \ref{fig:tracer_hist}). In the right panel the opposite effect is observed. The majority of the contamination
 now comes from \emph{smaller distances} where the tilt angle is \emph{smaller}, which in turns biases the mean observed tilt angle 
towards \emph{lower} values. This effect is observed for bins at distances large enough that there are few more distant stars left due 
to the survey's magnitude limit, and so it is more likely that stars from inner regions are scattered out to the $r_{obs}$ bin, as a 
consequence of the skewed distribution in distances that results from the computation from the reciprocal of the parallax \citep[see][for a detailed discussion]{brown05}.

The combination of these effects is what gives rise to the systematic trends observed in Figure \ref{fig:tilt_vs_r_mlimsample}, 
which affects GC3 results more dramatically causing it
 to systematically overestimate the tilt angle by $\sim2 \degr$ for most distances in the RC and A star samples. Although mGC3
 and nGC3 results are affected as well, Figure \ref{fig:tilt_vs_r_mlimsample} shows the key role played by the use of kinematical
 information for the recovery of the tilt angle using these methods is far less biased than with the purely position GC3. This
 comes from the fact that, although stars at different (true) distances are scattered into a particular observed distance bin, 
their kinematics are not consistent with the one corresponding to the position where they are observed. The mGC3 and nGC3
 velocity criteria naturally prevent these contaminant stars from contributing to a biased tilt angle.

\begin{figure}
\begin{center}
\includegraphics[width=65mm]{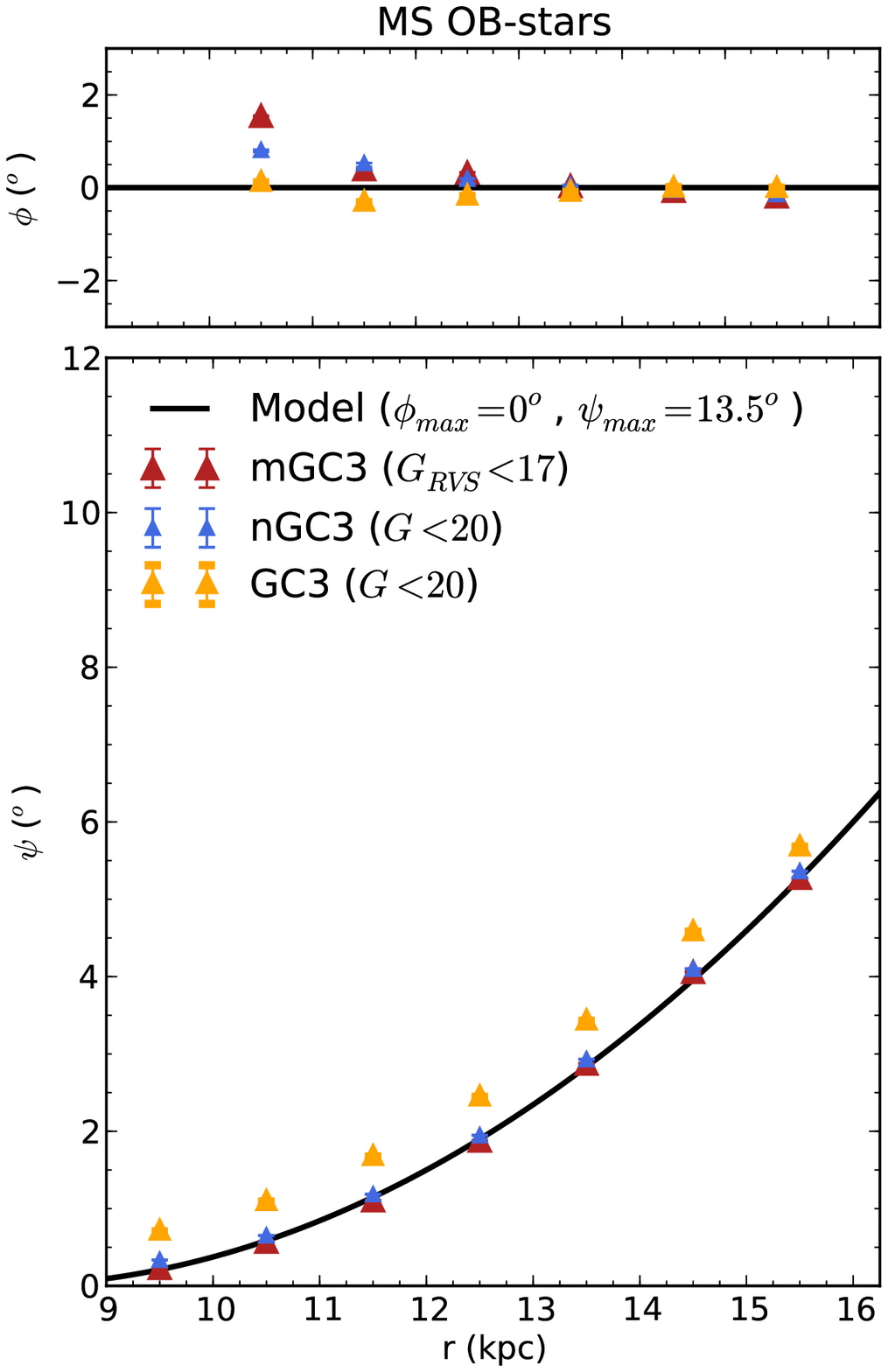}
 \caption{Tilt $\psi$ and twist $\phi$ angles versus $r$ for the Magnitude Limited sample of OB stars warped with UWH model.}
 \label{fig:tilt_vs_r_halfamp}                              
\end{center}
\end{figure}

We also explored the recovery of the parameters of the warp model for a case where the warp amplitude is significantly reduced
 ($\psi_{max}=13.5\degr$, UWH warp model)  
using a Magnitude Limited OB star sample. Results are presented in Figure \ref{fig:tilt_vs_r_halfamp}. The twist angle recovery is 
again very good and for the tilt angle, the behaviour is similar as that observed for this sample using our fiducial model (Figure 
\ref{fig:tilt_vs_r_mlimsample}, right panels). The best results for recovering the tilt angle are again obtained with mGC3.

\subsubsection{The Clean samples}
\label{subsec:ResCleanSamples}

\begin{figure*}
\includegraphics[width=150mm]{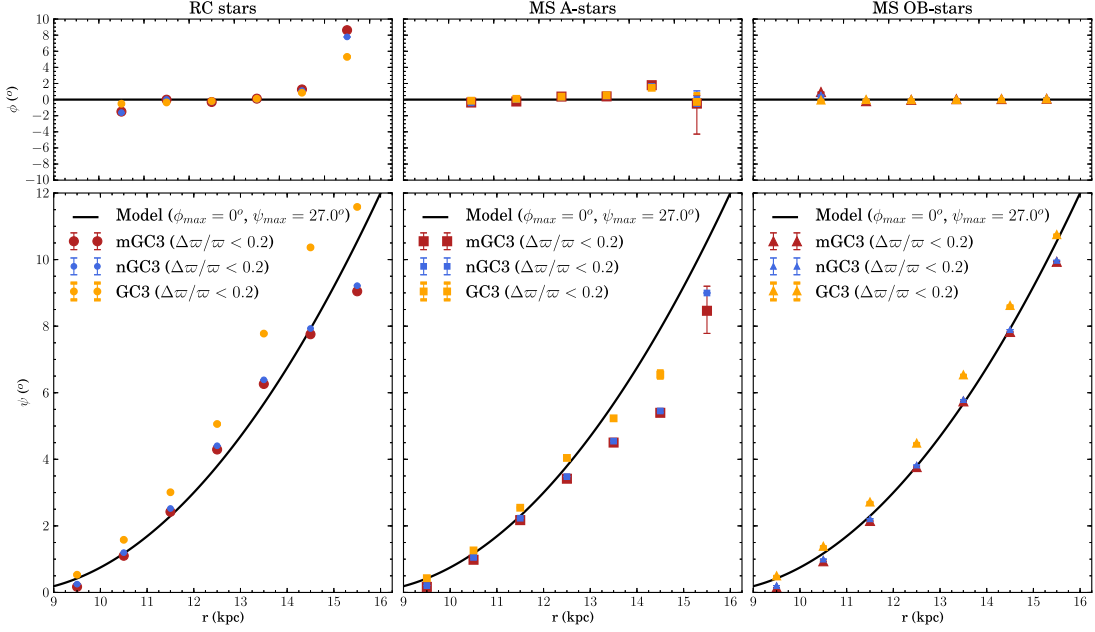}
 \caption{Tilt $\psi$ and twist $\phi$ angles versus Galactocentric (spherical) radius $r$ for the Clean sample ($\Delta \varpi/\varpi<0.2$).}
 \label{fig:tilt_vs_r_cleansample}
\end{figure*}

We now consider the \emph{Clean sample} that include only stars with parallax errors smaller than 20$\%$. 
Results are presented in Figure \ref{fig:tilt_vs_r_cleansample}.
We see that the recovery of the tilt angle is again excellent when using OB stars. Results for RC and A stars
are accurate for distances up to $\sim15$ kpc and $\sim13$ kpc, respectively. 
 The kinematical methods mGC3 and nGC3 again give the better results at all distances.
 The trends and biases explained before are still present, but their effect for the RC stars is reduced because the magnitude of the 
 errors is now smaller and with these being so numerous, sample size is not significantly compromised by the parallax error cut. 
 This is not the case, however, for the A star sample which is significantly reduced by the parallax cut (see green histograms in Figure \ref{fig:tracer_hist}), 
 for which the skewed distance bias starts dominating at shorter distances around $r\sim13$ kpc.
 
\begin{figure*}
\begin{center}
\includegraphics[width=180mm]{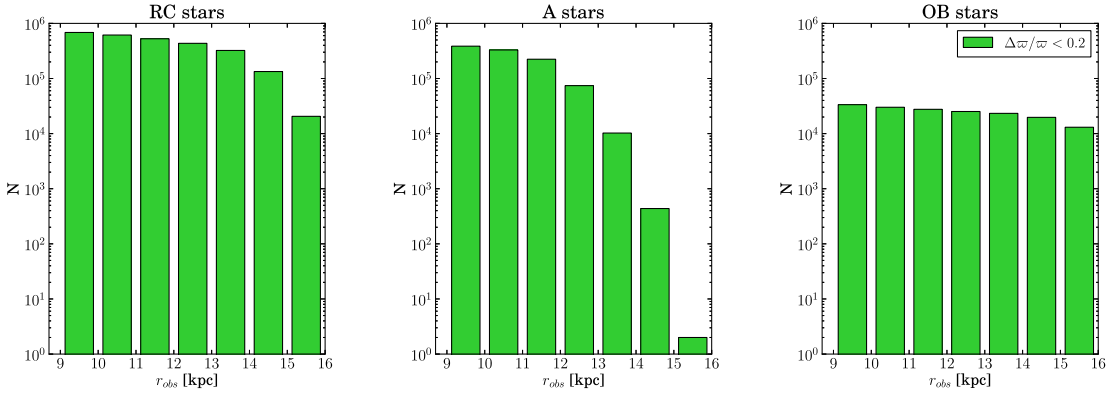}
 \caption{Histograms of the number of stars in \emph{observed} Galactocentric (spherical) radius bins of 1 kpc for the Clean samples 
($\Delta \varpi / \varpi <0.2$) of RC stars (\emph{left})
 A stars (\emph{middle}) and OB stars (\emph{right}).}
 \label{fig:hist_ob}                              
\end{center}
\end{figure*}

 Ideally one would like to have a criterion that would allow us to identify at which distance the results from the Clean sample
 start being significantly affected by this bias. For mGC3 and nGC3 we propose that this criterion can be defined
 empirically, looking at the number of stars in each observed distance bin for the Clean samples shown in Figure \ref{fig:hist_ob} . 
 This Figure shows that the distance for which the number of stars has decreased
down to $\lesssim10\%$ of the total stars in the innermost bin ($9<r_{obs}<10$ kpc), roughly coincides with the distance 
at which the bias in the tilt angle starts to dominate. \emph{Therefore, we can use this criterion as a rule of thumb to identify
 the distance up to which results from nGC3 and mGC3 methods can be trusted. } Applying it to the Clean sample results shown in Figure 
\ref{fig:tilt_vs_r_cleansample}, we can see that A stars at $r_{obs}>13$ kpc and RC stars at $r_{obs}>15$ kpc should be discarded.

%-----------------------------------------------------------------------------------------------
\subsection{Results for a ``twisted warp'' sample}
\label{subsec:ResTwistSample}
\begin{figure*}
\begin{center}
\includegraphics[width=160mm]{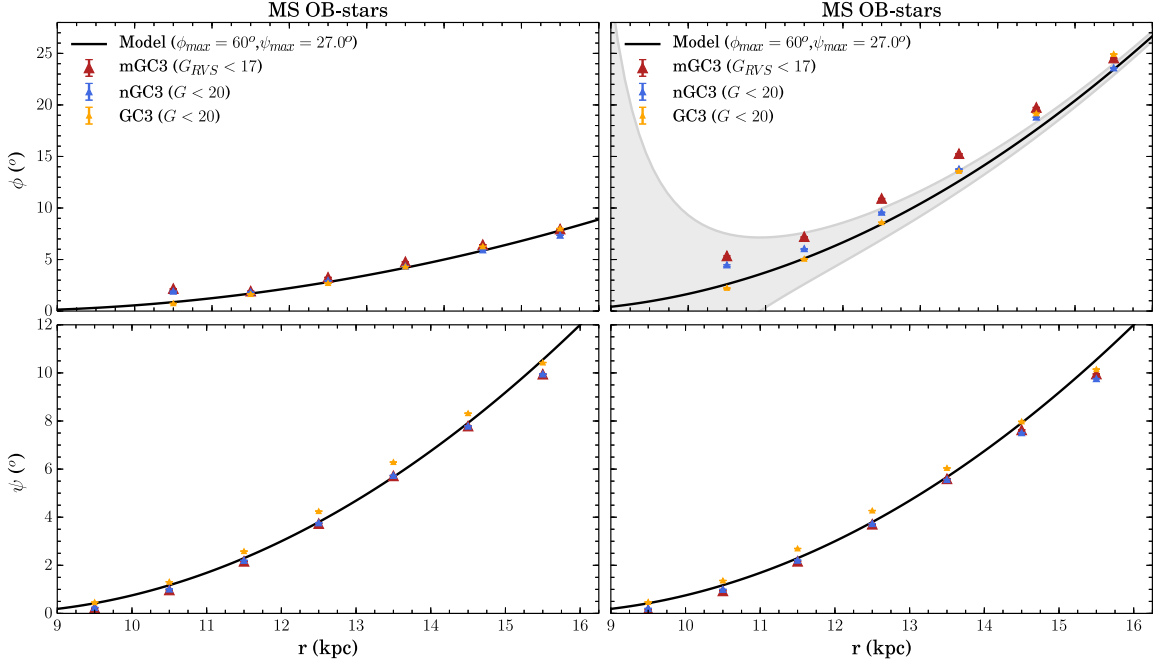}
 \caption{Tilt $\psi$ and twist $\phi$ angles versus $r$ for the Clean sample of OB stars, for two different twist models: TW1 
  (\emph{left}) and  TW2 (\emph{right}).  The shaded region in the upper right panel represents a  difference in twist angle (azimuthal coordinate) corresponding to an angular discrepancy of $\pm 0.1\degr$ with the model (see text). }
 \label{fig:torsion_tilt_twist_vs_r}
\end{center}
\end{figure*}

In this section we apply the same procedure to the Clean sample of OB stars, but now using two warp models \textit{including} twisting
 and with the same warp amplitude as our fiducial model (see Sec. \ref{subsec:twist}).  
Figure \ref{fig:torsion_tilt_twist_vs_r} shows the results for two different models: TW1 ($\phi _{max}=20^\circ$)
 (\emph{left} panels) and TW2 ($\phi _{max}=60^\circ$) (\emph{right} panels). The performance of all three methods is very good.
 The twist angle is 
recovered to within $<2\degr$, for nearly all distances, with the largest deviation still being no more than $\sim3\degr$.

 For the twist angle, we must consider that a given deviation becomes less significative as the tilt angle diminishes:  Since meridian lines in a polar grid converge toward the poles, close to them (in our case, small tilt angles), a given difference in twist angle translates into ever smaller angular deviations in the pole count maps (inversely proportional to sine of the tilt angle).  This is shown in the upper right panel of Figure \ref{fig:torsion_tilt_twist_vs_r}, where the shaded band represents a $\pm 0.1\degr$ variation in angular deviation with respect to the model. Note that this is not the same as a $\pm 0.1\degr$ variation in twist angle. Thus, the error in the recovered twist angle seen for the RC sample in the outermost bin in Figure \ref{fig:tilt_vs_r_cleansample}, is more significative than any of those seen in Figure \ref{fig:torsion_tilt_twist_vs_r}, which turns out to be very small in terms of actual angular deviation. 
%-----------------------------------------------------------------------------------------------
%====================================================
\section{Conclusions and expectations}
\label{sec:conclusions}

In this paper we test the capability of a family of Great Circle Cell Counts (GC3) methods to identify and characterize the warping of 
the stellar disc of our Galaxy in the Gaia era.
 These methods can work with samples for which full six-dimensional phase-space information
is provided \citep[mGC3 method, introduced by][]{mateu11}; samples for which radial velocity is lacking (nGC3
method, newly developed here); or samples having only positional information \citep*[GC3 method, firstly introduced by][]{johnston1996}.

We developed an analytical expressions for the force field of a warped Miyamoto-Nagai potential. Starting from 
the Galactic axisymmetric potential model of A$\&$S, we distort the potential according to two different warp
 models: 1) a model with a straight line of nodes and  2) a model with twisted line of nodes.
  Using a set of test particles that are relaxed in the A$\&$S potential, we warp the disc potential adiabatically, allowing the
 particles to follow the bended potential and not be left behind. In some cases a twist is introduced additionally through a 
purely geometric transformation of the particle's phase-space coordinates.The kinematic distribution of our synthetic samples mimic 
three different tracer populations: OB, A and Red Clump (RC) stars. 

The modified Great Circle Cell Counts method (mGC3) assume stars in a fixed Galactocentric ring are confined to a great circle band,
 with their Galactocentric position and velocity vector perpendicular to the normal vector which defines this particular great circle.
 The peak of the distribution in the pole count map, i.e. in the map of the number of stars associated to each great circle cell, is then
 identified using a Bayesian fitting procedure, which results in the identification of the tilt and twist angles
 of the warp and their corresponding confidence intervals. 
 
Considering the spatial distribution from the new Besan\c{c}on 
Galaxy Model and using the 3D extinction map
of \cite{drimmel03}, we have generated realistic mock catalogues of OB, A and RC stars where a very complete model
of Gaia observables and their expected errors are included. We have tested our methods and found their range of applicability,
identifying the main sources that limit them. We found that the introduction of the kinematic information in the methods (mGC3 and nGC3) 
 improves
  the recovery of the tilt angle to discrepancies less than $\sim0.75\degr$ for most of the cases, whereas using only positional information (GC3 method) 
the tilt angle recovery is systematically overestimated by $\sim2\degr$. Although seemingly small, for Galactocentric distances $r\lesssim12$ kpc,
 where the tilt angle is expected to be quite small, this $2\degr$ systematic trend represents an error of larger than $100\%$ in the tilt angle.
 We have been able to recognize the biases in the results introduced both
 by the fact that Gaia provides non-symmetric
 errors in trigonometric distances and that we are working in an apparent magnitude limited sample. The OB and RC stars samples
 are good warp tracers, whereas the A stars sample is not quite
 up to the task for Galactocentric distances larger than $\sim12$ kpc,
 mainly due to their fainter intrinsic luminosity. Using data with good astrometric quality (relative parallax accuracy of 20$\%$ or
 better), we obtain remarkably good accuracy recovering the tilt angle for all three tracers, provided we have enough stars in the
 Galactocentric
radius bins. We propose an empirical criterion to identify at which distance the nGC3 and mGC3 results from the Clean 
sample (the sample with relative parallax accuracy smaller than 20$\%$) start being significantly affected by biases.
 According to this rule of thumb, we should discard the Galactocentric radius 
bins for which the number of stars has decreased down to $\lesssim10\%$ of the total stars in the innermost bin ($9<r_{obs}<10$ kpc). 
 Using the Clean sample of OB stars warped with the twisted warp model, the twist angle is recovered to within $<3\degr$ for all distances.
 It is worth noting that throughout this paper we have used trigonometric parallaxes in the computation of pole counts
 with all three methods (Sec. \ref{sec:warpdetect}). For standard-candle tracers such as RC stars, or others not explored in this work such as RR Lyrae
 stars or Cepheids, an even better performance could be achieved with the use of photometric parallaxes for the faintest stars with
 large trigonometric parallax errors ($\Delta \varpi/\varpi>20\%$). By comparing the different variants of the methods, the power of exploiting kinematical 
information becomes apparent.
 
 In this work we have developed a first and simplified kinematic model for our Galactic warp. The simplicity of the model has
 allowed us to evaluate the efficacy and limitations of the use of Gaia data to characterize the warp. These limitations have
 been fully explored and quantified. In an upcoming paper we will present more complex and realistic
 developments that is generating lopsided warp models that will require a complex polar count maps analysis.
 From the work done so far, we expect that the Gaia database, together with
 the methods presented here, will be a very powerful combination to characterize the warp of the stellar disc of our Galaxy. 
%______________________________________________________________
\section*{Acknowledgments}
We would like to thank the anonymous referee for the comments on the manuscript. This work was carried out through the Gaia Research for European Astronomy Training (GREAT-
ITN) network funding from the European Union Seventh Framework Programme,  
under grant agreement $n\degr$264895, and 
and the MINECO (Spanish Ministry of Science and Economy) - FEDER
through grants  AYA2012-39551-C02-01 and CONSOLIDER CSD2007- 00050,
ESP2013-48318-C2-1-R and CONSOLIDER CSD2007-00050.
 LA acknowledges support from GREAT-ITN and the hospitality of the astronomy group at the University of Barcelona, where he spent
 a sabbatical year and this work got started. CM acknowledges the support of the postdoctoral Fellowship of DGAPA-UNAM, Mexico.
 We also acknowledge the Gaia Project Scientist Support Team and the Gaia Data
Processing and Analysis Consortium (DPAC) for providing the PyGaia toolkit. Simulations were carried out
using Atai, a high performance cluster, at IA-UNAM.
\bibliographystyle{mn2e}

\def\apj{ApJ}
\def\apjl{ApJ}
\def\aj{AJ}
\def\mnras{MNRAS}
\def\aa{A\&A}
\def\nat{nat}
\def\araa{ARA\&A}
\def\aap{A\&A}
\def\rmxaa{RMAA}
\def\pasp{PASP}
\def\pasj{PASJ}
\def\aaps{AAS}
\def\apjs{ApJS}
\def\bain{Bulletin of the Astronomical Institutes of the Netherlands}

\bibliography{WarpPaper_subm}
\IfFileExists{\jobname.bbl}{}
{\typeout{}
\typeout{****************************************************}
\typeout{****************************************************}
\typeout{** Please run "bibtex \jobname" to optain}
\typeout{** the bibliography and then re-run LaTeX}
\typeout{** twice to fix the references!}
\typeout{****************************************************}
\typeout{****************************************************}
\typeout{}
}

\appendix

\section[]{Calculation of the warped forces}
\subsection{Forces of the warped Miyamoto--Nagai disc}

To calculate the force field of the warped Miyamoto-Nagai
potential we proceed as follows: Let $x^{\prime}_i$ be a cartesian
coordinate system and $x_i$ the coordinate system we obtain
by applying a warping transformation $W$:
\begin{equation}
W: x^{\prime}_i \rightarrow x_i,
\end{equation}
e.g. the lines of fixed coordinate $x_i$ look warped in the
coordinate system $x^{\prime}_i$.
Now, we want to compute the force field in the coordinate
system $x^{\prime}$. In the coordinate system $x$ the force field
is just the one produced by the original flat disc (because
these coordinate are warped too). Since the force is a
covariant vector, it transforms as:
\begin{eqnarray}\label{warp_force}
F^{\prime}_j(x^{\prime}) & =& - {{\partial \Phi(x^{\prime})}\over{\partial x^{\prime}_j}}  \nonumber \\
&=& - \sum_{i=1}^3 {{\partial \Phi(x[x^\prime])}\over{\partial x_i}} {{\partial x_i}\over{\partial x^{\prime}_j}}\nonumber \\
&=&  \sum_{i=1}^3 F_i(x[x^\prime]) {{\partial x_i}\over{\partial x^{\prime}_j}}
\end{eqnarray}

So this is our force transformation: $F(x)$ is the force field
of a flat disc and $F^{\prime}(x^{\prime})$ is the corresponding warped field
under the $W$ transformation. The force field
of the flat Miyamoto-Nagai potential is as follows:
\begin{equation}\label{f1}
F_{x_1}=\frac{x_1}{(x_1^ { 2}+x_2^{ 2} + (a+ \sqrt{x_3^{ 2}+ b^2})^2)^{\frac{3}{2}}}
\end{equation}

\begin{equation}
F_{x_2}=\frac{x_2}{(x_1^ { 2}+x_2^{2} + (a+ \sqrt{x_3^{ 2}+ b^2})^2)^{\frac{3}{2}}}
\end{equation}

\begin{equation}\label{f3}
F_{x_3}=\frac{x_3(a+\sqrt{x_3^{ 2} + b^2})}{\sqrt{x_3^{ 2} + b^2}(x_1^ { 2}+x_2^{2} + (a+ \sqrt{x_3^{ 2}+ b^2})^2)^{\frac{3}{2}}}
\end{equation}
All we need to do now is to
obtain the elements of the Jacobian matrix of $W$.

The warp is accomplished by a simple rotation around the
$X$-axis, which constitutes our line of nodes:
\begin{equation}\label{one}
 \begin{pmatrix}
x_1 \\
x_2\\ 
x_3\\
\end{pmatrix}= 
\begin{pmatrix}
1 & 0 &  0 \\
0               & cos (\psi) &  -sin(\psi)                \\ 
0   & sin(\psi) &  cos(\psi)     \\
\end{pmatrix}
\begin{pmatrix}
x_1^{\prime}  \\
x_2^\prime  \\ 
x_3^\prime \\
\end{pmatrix}
\end{equation}
Where $\psi (r^\prime) = \psi_2 (\frac{r^\prime-r_1}{r_2 - r_1})^\alpha$ . Note that $r^\prime$ is the spherical Galactocentric radius. 
The $\frac{\partial x_i}{ \partial x^\prime_j} $ terms can easily be calculated from Equations \ref{one}:
\begin{equation}\label{jacob1}
\frac{\partial x_1}{\partial x_1^\prime}  = 1
\end{equation}
\begin{eqnarray}
\frac{\partial x_2}{\partial x_1^\prime}  =  -x_2^{\prime}\, sin(\psi) \, \left(\frac{\partial \psi}{\partial x_1^{\prime}}\right) - x_3^{\prime}\,  cos(\psi) \, \left(\frac{\partial \psi}{\partial x_1^{\prime}}\right)
\end{eqnarray}
\begin{eqnarray}
\frac{\partial x_3}{\partial x_1^\prime}  =  x_2^{\prime}\, cos(\psi) \, \left(\frac{\partial \psi}{\partial x_1^{\prime}}\right) - x_3^{\prime} \, sin(\psi) \, \left(\frac{\partial \psi}{\partial x_1^{\prime}}\right)
\end{eqnarray}
\begin{equation}
\frac{\partial x_1}{\partial x_2^\prime}  =  0
\end{equation}
\begin{eqnarray}
  \frac{\partial x_2}{\partial x_2^\prime}  = cos(\psi) -x_2^{\prime}\, sin(\psi) \left(\frac{\partial \psi}{\partial x_2^{\prime}}\right) \nonumber \\
- x_3^{\prime} \, cos(\psi) \left(\frac{\partial \psi}{\partial x_2^{\prime}}\right)
\end{eqnarray}
\begin{eqnarray}
\frac{\partial x_3}{\partial x_2^\prime}  =  sin(\psi )+x_2^{\prime}\, cos(\psi) \left(\frac{\partial \psi}{\partial x_2^{\prime}}\right) \nonumber \\
- x_3^{\prime}\,  sin(\psi) \left(\frac{\partial \psi}{\partial x_2^{\prime}}\right)
\end{eqnarray}
\begin{equation}
\frac{\partial x_1}{\partial x_3^\prime}  = 0
\end{equation}
\begin{eqnarray}
\frac{\partial x_2}{\partial x_3^\prime} = -sin (\psi )-x_2^{\prime}\, sin(\psi) \left(\frac{\partial \psi}{\partial x_3^{\prime}}\right) \nonumber \\
- x_3^{\prime}\,cos(\psi) \left(\frac{\partial \psi}{\partial x_3^{\prime}}\right)
\end{eqnarray}
\begin{eqnarray}\label{jacob2}
\frac{\partial x_3}{\partial x_3^\prime}  =cos (\psi )-x_2^{\prime}\, cos(\psi) \left(\frac{\partial \psi}{\partial x_3^{\prime}}\right) \nonumber \\
- x_3^{\prime}\, sin(\psi) \left(\frac{\partial \psi}{\partial x_3^{\prime}}\right)
\end{eqnarray}

Where:
\begin{equation}
\frac{\partial \psi(r^\prime)}{\partial x_1^{\prime}}= \frac{\alpha \, \, \psi _{2}\,\, x_1^\prime}{r^\prime(r_2 -r_1)}\, \,\left(\frac{r^\prime-r_1}{r_2 - r_1}\right)^{\alpha -1}
\end{equation}
\begin{equation}
\frac{\partial \psi(r^\prime)}{\partial x_2^{\prime}}= \frac{\alpha \, \, \psi _{2}\,\, x_2^\prime}{r^\prime(r_2 -r_1)}\, \,\left(\frac{r^\prime-r_1}{r_2 - r_1}\right)^{\alpha -1}
\end{equation}
\begin{equation}
\frac{\partial \psi(r^\prime)}{\partial x_3^{\prime}}= \frac{\alpha \, \, \psi _{2}\,\, x_3^\prime}{r^\prime(r_2 -r_1)}\, \,\left(\frac{r^\prime-r_1}{r_2 - r_1}\right)^{\alpha -1}
\end{equation}
Now using these elements of the Jacobian matrix, the warped force field 
in cartesian coordinates can easily be calculated using Equation \ref{warp_force}.
\bsp
\label{lastpage}

\end{document}